\documentclass[prb,twocolumn,amsmath,amssymb]{revtex4}

\usepackage{graphicx}
\usepackage{epstopdf}
\usepackage{dcolumn}
\usepackage{bm}
\usepackage{color}

\usepackage{epsfig,float,afterpage,amssymb,wrapfig,psfrag}
\usepackage[caption=false]{subfig}

\DeclareMathAlphabet{\bi}{OML}{cmm}{b}{it}

\DeclareMathOperator{\deter}{det}

\begin{document}
\title{Bose-Fermi Anderson Model with SU(2) Symmetry: Continuous-Time Quantum Monte Carlo Study}

\author{Ang Cai}
\affiliation{Department of Physics and Astronomy, Rice Center for Quantum Materials, Rice University,
Houston, Texas, 77005, USA}

\author{Qimiao Si}
\affiliation{Department of Physics and Astronomy, Rice Center for Quantum Materials, Rice University,
Houston, Texas, 77005, USA}
\date{\today}
\begin{abstract}
In quantum critical heavy fermion systems, local moments are coupled to both collective spin fluctuations 
and conduction electrons. As such, the Bose-Fermi Kondo model, describing the coupling of a local moment
to both a bosonic and a fermionic bath, has been of extensive interest. For the model in the presence 
of SU(2) spin rotational symmetry, questions have been raised about its phase diagram.
Here we develop a version of continuous-time Quantum Monte Carlo (CT-QMC) 
method suitable for addressing this issue; this procedure
 can reach sufficiently low temperatures
while preserving the SU(2) symmetry.
Using this method for the Bose-Fermi Anderson model,
we clarify the renormalization-group fixed points and the phase diagram for the case 
with a constant fermionic-bath density of states and
  a power-law bosonic-bath spectral function $\rho_{b}(\omega) \propto \omega^{s}$ ($0<s<1$).
We find two types of Kondo destruction QCP, depending on the power-law exponent $s$ in the bosonic bath spectrum.
For $s^{*}<s<1$, both types of QCPs exist and, in the parameter regime accessible by an analytical
$\epsilon$-expansion renormalization-group calculation (here $\epsilon=1-s$), the  CT-QMC
result is fully consistent 
with prior predictions by the latter method. 
For $s<s^{*}$, there is only one type of QCP.
At both type of Kondo destruction QCPs, 
we find that the exponent of the local spin susceptibility 
$\eta$ obeys the relation $\eta=\epsilon$, which
 has important implications
for Kondo destruction QCP in the Kondo lattice problem.
\end{abstract}

\maketitle
\section{Introduction}
Heavy fermion systems serve as a prototype system to study quantum criticality 
\cite{si2010heavy,Coleman-Nature}.
 Experimental discoveries in various heavy fermion compounds 
open up the opportunity to explore
beyond-Landu type quantum critical points (QCP) 
in the context of antiferromagnetic Kondo lattice systems. One prominent example is the Kondo destruction QCP 
\cite{si2001.nature,Colemanetal,senthil2004a},
where the phase transition 
at zero temperature not only involves the 
magnetic order parameter, but also the localization to delocalization transition of the 4f electrons constituting  the local moments. Some of the hallmarks of Kondo destruction type QCP involves $\omega/T$ scaling of the dynamical spin susceptibility as seen from inelastic neutron scattering, jump of the fermi surface volume from magnetotransport and quantum oscillation measurement \cite{si2013quantum}.
Such properties 
 are inconsistent with predictions from the traditional spin-density-wave type QCP \cite{hertz1976quantum,millis1993effect,moriya2012spin}.

One of the simplest models that contain a Kondo destruction type QCP is the Bose-Fermi Kondo model (BFKM)
\cite{si2014kondo}. It arises in the context of understanding the competition between Kondo effect and 
magnetic fluctuations in Kondo lattice model using extended dynamical mean field theory (EDMFT) \cite{si2001.nature,si2003.prb}. It describes a local moment coupled to both itinerant electrons 
as well as free bosons, which are usually referred to as fermionic bath and bosonic bath. 
Typically the fermionic bath will assume a constant density of states, and the bosonic bath has a sub-ohmic spectrum:
its density of states at low frequencies ($\omega$) have a power-law form,
$\rho_{b}(\omega) \propto \omega^{s}$ with $s<1$. It characterized the softened spectrum of the magnons 
near the magnetic QCP,  which competes with the conduction electrons in their couplings to the local moment
and causes the suppression of the Kondo effect.

This model is first treated with $\epsilon$-expansion renormalization group (RG) method, 
using $\epsilon=1-s$ as a small parameter \cite{si1996kosterlitz,smith1999,sengupta2000,si2001.nature,si2003.prb,zhu2002,zarand2002}. 
It turns out the fixed point structure will depend on the symmetry of the spin boson coupling: for the SU(2) and XY symmetric cases, it has a Kondo screened stable fixed point (K) at strong coupling, a bosonic bath dominated stable fixed point (L) at intermediate coupling (so called critical phase), and an unstable critical point (C) describing the quantum phase transition. Both L and C can be accessed by the $\epsilon$-expansion; for the Ising anisotropic case, on the other hand, the critical phase controlled by L is unstable and is replaced by the local moment fixed point (L$^{\prime}$) at strong coupling. In all three cases, it is predicted that at the critical point where the Kondo effect is critically destroyed, the local spin correlation function will behave as $\chi_{spin}(\tau) \sim (1/\tau)^{\eta}$, with an exact relation $\eta=\epsilon$ \cite{zhu2002,zarand2002}. This has important implications for the EDMFT calculation of the Kondo lattice problem. For two dimensional magnetic fluctuations, it predicts a Kondo destruction QCP solution, provided that the relation $\eta=\epsilon$ will remain valid at $\epsilon \rightarrow 1^{-}$.

The numerical calculations of the Bose-Fermi Kondo model and the closely related Bose-Fermi Anderson model (BFAM)
include treating it either as a standalone model using numerical renormalization group (NRG) \cite{glossop2005,glossop2007prb} and continuous-time quantum Monte Carlo (CT-QMC) \cite{pixley2011,pixley2013,otsuki}, or as an effective model under EDMFT \cite{grempel2003,jxzhu2003,glossop2007prl,jxzhu2007}.
Our focus in this work is on the CT-QMC method,
from which a seeming controversy existed for the 
SU(2) symmetric BFAM~\cite{otsuki}:
for $s=0.2$, it was shown that 
the Kondo-destruction phase has the local-moment character instead of being critical;
 in the temperature dependence of the local spin susceptibility in this Kondo-destruction phase,
 it was  found $\chi^{spin} \sim 1/T$ instead of the $\chi^{spin} \sim 1/T^{s}$ behavior 
 predicted by $\epsilon$-expansion RG\cite{zhu2002,zarand2002} 
 for fixed point L.

\begin{figure}[!htb]
\captionsetup[subfigure]{labelformat=empty}
  \centering
  \mbox{\includegraphics[width=0.75\columnwidth]{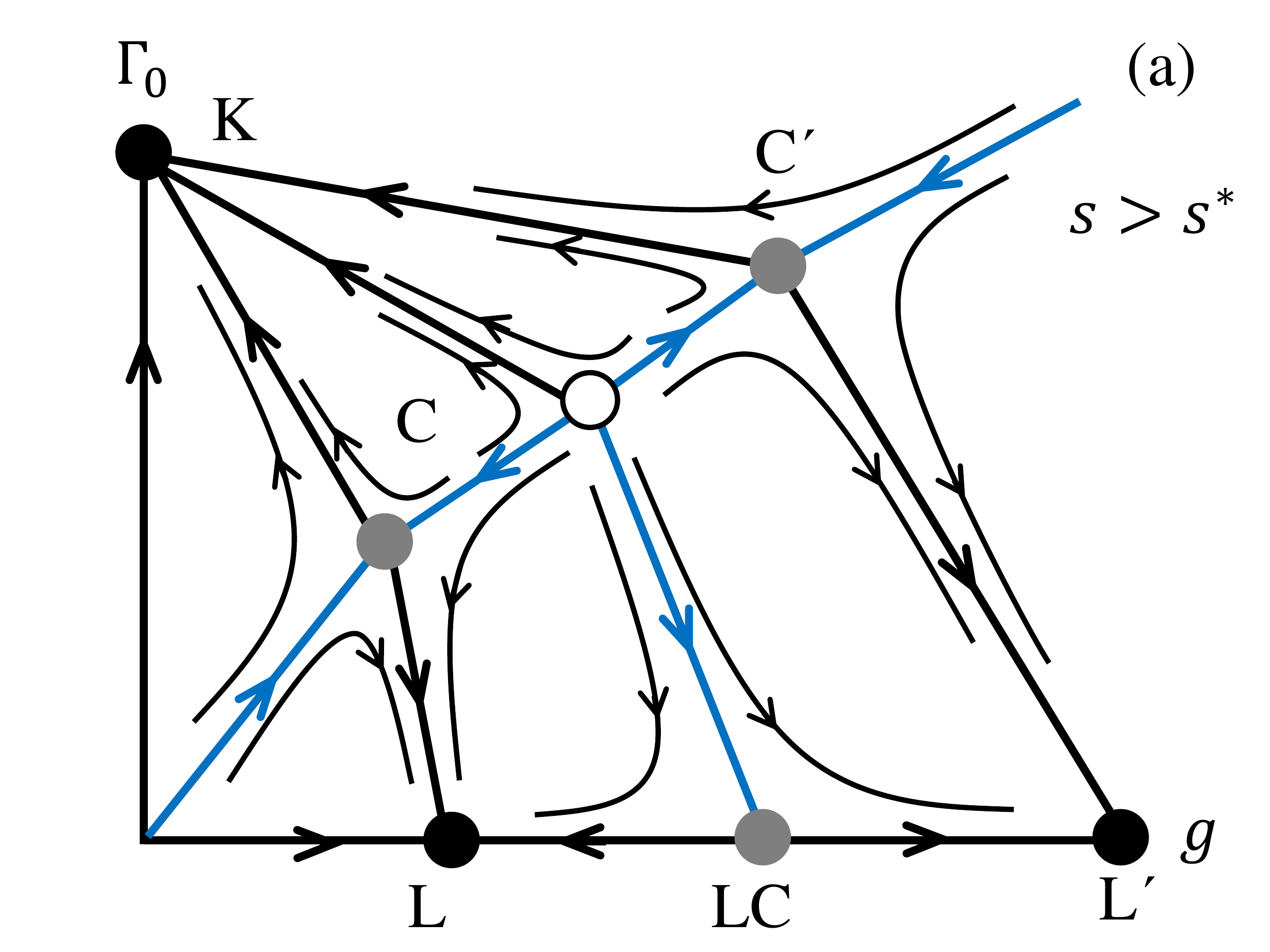}}    
  \mbox{\includegraphics[width=0.75\columnwidth]{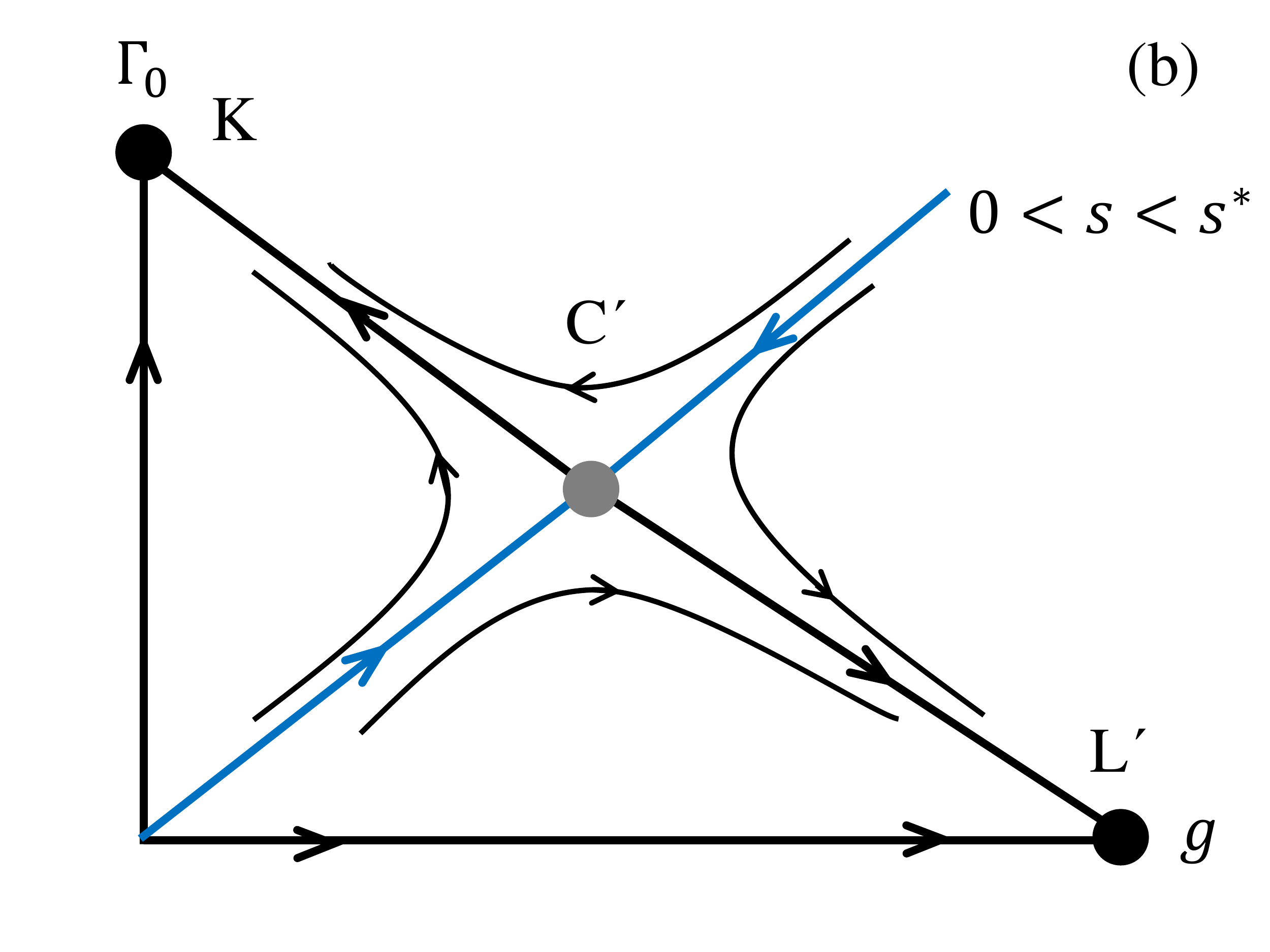}}                            
\caption{RG flow of SU(2) BFAM suggested by our CT-QMC results. 
Filled black (gray) dots represent stable (unstable) fixed points. Blue lines denote separatrix between different stable phases. (a) $s^{*}<s<1$: 
There are two stable fixed points L and L$^{\prime}$, one unstable fixed point LC along $\Gamma_{0}=0$ axis, and one stable Kondo fixed points K along $g=0$ axis. C and C$^{\prime}$ are two unstable fixed points associated with Kondo destruction towards fixed points L and L$^{\prime}$ . (b) $0<s<s^{*}$: fixed point L  disappears, 
leaving only one unstable fixed point C$^{\prime}$ between Kondo and stable fixed point L$^{\prime}$ . 
We have estimated $s{*} \simeq 0.47$, as shown in Fig.~\ref{phase_b}.
}
\label{fig:rg}
\end{figure}

To resolve this seeming inconsistency,
we start with the observation that, if $s$ is close to $1$,
 the CT-QMC result must be consistent with that of the 
 $\epsilon$-expansion RG in the range of coupling constants 
accessed by this expansion (again $\epsilon=1-s$).
 To make progress,
in this article we 
develop the CT-QMC procedure for the BFAM such that it can reach sufficiently low temperatures
while preserving the SU(2) symmetry.
Using this procedure,
 we  carry out
 a comprehensive study of the SU(2) BFAM
 for $s$ ranging from close to $0$ to close to $1$.
 We 
 study 
a variety of
 observables in order to identify all the QCPs between different phases, 
 combined with detailed finite size scaling analysis to extract critical exponents. 
 
 Our analysis shows that the $\epsilon$-expansion~\cite{zhu2002,zarand2002}
  and CT-QMC results are fully compatible with each other.
 Our results are summarized by the RG-flow diagrams of figure~\ref{fig:rg}. 
 For the $s>s^{*}$ regime, we identify i) the critical point C separating the Kondo screened phase and critical phase, as predicted from $\epsilon$-expansion RG for the coupling constants accessible by the latter method; and ii)
 a separate critical point C$^{\prime}$ and stable fixed point L$^{\prime}$, which occurs for larger values 
 of the bosonic-Kondo coupling $g$.
 For $s<s^{*}$, there exists only a type ii) quantum phase transition~\cite{otsuki}.
We also determine the correlation length exponent $\nu $.
Additionally,
we find another unstable fixed point LC that controls the transition from fixed point L and fixed point L$^{\prime}$. 
Finally, we quantitatively estimate $s^{*}$ and conclude that the result at $s=0.2$ falls outside the 
regime that is controlled by 
the $\epsilon$-expansion.

The remainder of the paper is organized as follows. In Sec.~\ref{model} we introduce the SU(2) 
Bose-Fermi Anderson model, and give an overview of the CT-QMC method as well as the physical 
quantities we will investigate in this work. We will present the numerical results in Sec.~\ref{numerical}. 
We will start with a detailed study for the $s=0.6$ case in Sec.~\ref{numerical.A}, 
followed by the $s=0.2$ case in Sec.~\ref{numerical.B},
before carrying through the analysis that leads to an estimate for the value of $s^{*}$ in Sec.~\ref{numerical.C}. 
We will discus the implication of our results in Sec.~\ref{discussion} and conclude the article 
 in Sec.~\ref{conclusions}.

\section{Model and Method}
\label{model}
The Hamiltonian for the SU(2) symmetric BFAM reads,
\begin{equation}
H=H_{c}+H_{b}+H_{d}+H_{g}+H_{V},
\label{eq:bfkm}
\end{equation}
where $H_{c}$ and $H_{b}$ describes the bosonic and fermionic bath part, respectively,
\begin{eqnarray}
H_{c}=\sum_{k,\sigma} \epsilon_{k} c_{k,\sigma}^{\dagger} c_{k,\sigma} , H_{b}=\sum_{\alpha} H_{b}^{\alpha}=\sum_{p,\alpha}\omega_{q}  {\phi_{p}^{\alpha}}^{\dagger} \phi_{p}^{\alpha}.
\end{eqnarray}
$H_{d}$ contains the local electron part,
\begin{equation}
H_{d}=\sum_{\sigma} \epsilon_{d} d_{\sigma}^{\dagger} d_{\sigma} + U d^{\dagger}_{\uparrow} d_{\uparrow} d^{\dagger}_{\downarrow} d_{\downarrow}  \nonumber .
\end{equation}
$H_{V}$ and $H_{g}$ couples the local orbital to the bosonic and fermionic bath,
\begin{eqnarray}
H_{V}=\sum_{k,\sigma}  V d_{\sigma}^{\dagger} c_{k,\sigma} + h.c. , 
H_{g}=\sum_{p,\alpha} g S_{\alpha} ( {\phi^{\alpha}_{p}}^{\dagger} + \phi_{-p}^{\alpha} )
\end{eqnarray}
where the summation over $\alpha$ runs through $x$,$y$,$z$, $S_{\alpha}=d^{\dagger}_{\sigma} \tau^{\alpha}_{\sigma \sigma^{\prime} } d_{ \sigma^{\prime} }$, and $\tau^{\alpha}_{\sigma \sigma^{\prime} }$ being the three components Pauli matrices.

The properties of the fermionic and bosonic bath are specified by their density of states. For the fermionic bath, we choose a constant density of states,
\begin{equation}
\rho_{F}(\epsilon)=\sum_{k} \delta(\epsilon-\epsilon_{k}) = \rho_{0} \Theta (|D-\epsilon|),
\end{equation}
which leads to a hybridization function $\Gamma(\epsilon)= \Gamma_{0} \Theta(|D-\epsilon|)$, with $\Gamma_{0}=\pi\rho_{0}V^{2}$. 

Unless specified otherwise, the density of states for the sub-Ohmic bosonic bath has an exponential cutoff, given by the following,
\begin{equation}
\rho_{b}(\omega)=\sum_{q} \delta(\omega-\omega_{q} ) = K_{0} \omega^{s} e^{-\omega/\Lambda} \Theta(\omega).
\end{equation}

Throughout the text we fix $D=1$, $\Lambda=1$, and stays at the particle-hole symmetric point $U=-2\epsilon_d=0.1$. 
The prefactor $\rho_{0}$ and $K_{0}$ in the density of states of the fermionic bath and bosonic bath are determined 
from the normalization condition $\int_{-D}^{D} \rho_{F}(\epsilon) d \epsilon = 1$ and $ \int_{0}^{\infty} \rho_{b}(\omega) 
d\omega =1 $. We will use either the amplitude of the hybridization function $\Gamma_{0}$ or the spin-boson 
coupling $g$ as our tuning parameter.

\subsection{Monte-Carlo procedure}
We will employ the CT-QMC algorithm, first introduced in reference~\cite{werner2006prl,werner2006prb} and then generalized 
to treat the BFAM in references~\cite{pixley2011,pixley2013,otsuki}.
We start with removing the $z$ component of the spin-boson coupling 
by employing a Firsov-Lang transformation $\tilde{H}=e^{S} 
H e^{-S}$ with $S=g S_{z} \sum_{p}\frac{1}{\omega_{p}} ( {\phi_{p}^{z}}^{\dagger} +\phi_{-p}^{z} )$ 
(similar to Ref.\,\onlinecite{werner2007}) and work with the transformed Hamiltonian 
 $\tilde{H}$,
\begin{eqnarray}
\tilde{H} &=& H_{c}+H_{b}+\tilde{H}_{d}+\tilde{H}_{V}+\tilde{H}_{g} \nonumber \\
\tilde{H}_{d}&=&\sum_{\sigma} \tilde{\epsilon} {d}_{\sigma}^{\dagger} d_{\sigma} + \tilde{U} {d}^{\dagger}_{\uparrow} {d}_{\uparrow} {d}^{\dagger}_{\downarrow} {d}_{\downarrow}  \nonumber \\
\tilde{H}_{V}&=& V \sum_{k,\sigma} \left({d}_{\sigma}^{\dagger} c_{k,\sigma} 
e^{\sum_{p} \frac{gs_{\sigma}}{\omega_{p}}({\phi^{z}_{p}}^{\dagger}-{\phi^{z}_{p}}) }+h.c.\right) \nonumber \\
\tilde{H}_{g}&=&\sum_{p} (g / \sqrt{2} )
\left(
{S}_{+} \phi_{p}^{-} e^{\sum_{p} \frac{g }{\omega_{p}}({\phi^{z}_{p}}^{\dagger}-{\phi^{z}_{p}}) }
\right.
\nonumber
\\&+&\left.
{S}_{-} \phi_{p}^{+} e^{-\sum_{p} \frac{g }{\omega_{p}}({\phi^{z}_{p}}^{\dagger}-{\phi^{z}_{p}}) }
\right),
\end{eqnarray} 
where we have defined the renormalized parameters $\tilde{\epsilon}_{d}=\epsilon_{d}-(g^{2}/4) \sum_{q} (1/\omega_{q})^{2}$, $\tilde{U}=U+(g^{2}/2) \sum_{q} (1/\omega_{q})^{2}$, $s_{\sigma}=\pm 1/2 $ for $\sigma=\uparrow/\downarrow$.
and recombined the $x$ and $y$ components of $S_{\alpha}$ and $\phi_{\alpha}$ into
$S_{+}=d^{\dagger}_{\uparrow} d_{\downarrow}$, $S_{-}=d^{\dagger}_{\downarrow} d_{\uparrow}$, $\phi_{p}^{\pm}=(1/\sqrt{2})\left(({\phi_{p}^{x}}^{\dagger}+\phi_{p}^{x})\pm i ({\phi_{p}^{y}}^{\dagger}+\phi_{p}^{y})\right)$. 
The partition function is constructed by expanding in the non-diagonal terms
\cite{werner2006prl,werner2006prb,pixley2011,pixley2013,otsuki},
 $\tilde{H}_{V}$ and $\tilde{H}_{g}$ 
under the interaction representation of $H_{0} \equiv H_{b}+H_{c}+\tilde{H}_{d}$. It has the following form
\cite{pixley2011,pixley2013,otsuki}:
\begin{eqnarray}
Z&=&Z_{0} \sum_{m}
\int \prod_{i=1}^{m} d\tau^{s}_{i} d\tau^{s^{\prime}}_{i}
\prod_{\sigma=\uparrow,\downarrow}\left( \int\prod_{i=1}^{n_{\sigma}}d\tau^{d\sigma}_{i}
d\tau^{d^{\prime}\sigma}_{i}
\right)
\nonumber \\
&&
w_{d}(\{\tau^{tot}\}_{n_{tot}} )
\prod_{\sigma=\uparrow,\downarrow}
w^{\sigma}_{c}(\{\tau^{d\sigma}\}_{n_{\sigma}},\{\tau^{d^{\prime}\sigma}\}_{n_{\sigma}})
\nonumber \\
&& 
w_{z}(\{\tau^{tot}\}_{n_{tot}} )
w_{p}(\{\tau^{s}\}_{m},\{\tau^{s^{\prime}}\}_{m}),
\end{eqnarray}
where $Z_{0}=Tr[e^{-\beta H_{c}}] Tr[e^{-\beta H^{z}_{B}}] Tr[e^{-\beta (H^{x}_{B}+H^{y}_{B})}] $ is the partition function of the bath, $\beta$ being the inverse temperature: $\beta=1/T$.
$\int \prod_{i=1}^{m} d\tau^{\alpha}_{i} d\tau^{\alpha^{\prime}}_{i}=\int_{0}^{\beta} d\tau^{\alpha}_{1}    \cdots  \int_{\tau^{\alpha}_{N-1}}^{\beta} d\tau^{\alpha}_{N}\int_{0}^{\beta} d\tau^{\alpha^{\prime}}_{1}   \cdots  \int_{\tau^{\alpha^{\prime}}_{N-1}}^{\beta} d\tau^{\alpha^{\prime}}_{N}$.
$\{\tau^{\alpha}\}_{n}$ denotes the set of imaginary time of all the operators of a given type $\alpha$ in the expansion: $\{\tau^{\alpha}\}_{n}=\{\tau^{\alpha}_{1},\tau^{\alpha}_{2}\dots,\tau^{\alpha}_{n}\}$. $\alpha \in \{s,s^{\prime},d\sigma, d\sigma^{\prime} \}$ represents $S_{+}$, $S_{-}$, $d^{\dagger}_{\sigma}$, or $d_{\sigma}$. $n=m$ or $n_{\sigma}$ denotes the number of pairs of $S_{+}, S_{-}$ or $d^{\dagger}_{\sigma}, d_{\sigma}$, also labeling the expansion order.
$\{\tau^{tot}\}_{n_{tot}}$ refers to all the $\{\tau^{\alpha}\}_{n}$ combined,
with $n_{tot}=2(\sum_{\sigma} n_{\sigma} + m)$. The integrand, or so-called weight, factorizes into multiple components. In the following we will present the form of each part explicitly.

$w_{d}(\{\tau^{tot}\}_{n_{tot}} )$ is the contribution from the local d electron part. It describes valence and spin fluctuations of the local orbitals,
\begin{eqnarray}
w_{d}&=&Tr[e^{-\beta \tilde{H}_{d} }T_{\tau}  S_{-}(\tau^{s^{\prime}}_{m}) S_{+}(\tau^{s}_{m}) \cdots S_{-}(\tau^{s^{\prime}}_{1}) S_{+}(\tau^{s}_{1})  \nonumber \\
&\times& \prod_{\sigma} d_{\sigma}(\tau^{d^{\prime}\sigma}_{n_{\sigma}}) d_{\sigma}^{\dagger}(\tau^{d\sigma }_{n_{\sigma}}) \cdots d_{\sigma}(\tau^{d^{\prime}\sigma}_{1}) d_{\sigma}^{\dagger}(\tau^{d\sigma }_{1})].
\end{eqnarray}
Here for a given operator $O$, $O(\tau)$ denotes the corresponding operator in the interaction representation $O(\tau)=e^{\tau H_{0} } O e^{-\tau H_{0}}$.

$w^{\sigma}_{c}(\{\tau^{d\sigma}\}_{n_{\sigma}},\{\tau^{d^{\prime}\sigma}\}_{n_{\sigma}})$ is the contribution from the conduction electron with spin index $\sigma$,
\begin{eqnarray}
w^{\sigma}_{c}&=& V^{2 n_{\sigma} } \left( \prod_{i=1}^{n_{\sigma} }  \sum_{k_{i},k^{{\prime}}_{i}}\right) Tr[T_{\tau} e^{-\beta H_{c} }  c^{\dagger}_{k_{n_{\sigma}},\sigma}(\tau^{d^{\prime}\sigma}_{n_{\sigma}})
\nonumber \\
&\times& c_{k^{\prime}_{n_{\sigma}},\sigma}(\tau^{d\sigma}_{n_{\sigma}}) \cdots c^{\dagger}_{k_{1},\sigma}(\tau^{d^{\prime}\sigma}_{1})
c_{k^{\prime}_{1},\sigma}(\tau^{d\sigma}_{1})]/Tr[e^{-\beta H_{c} }]
\nonumber \\
&=&\deter(F^{\sigma}).
\end{eqnarray}
It can be expressed as a determinant of matrix $F^{\sigma}$, whose matrix element is given by 
\begin{eqnarray}
F^{\sigma}_{ij}=\frac{-\sum_{k}V^{2} Tr[e^{-\beta H_{c} } T_{\tau} c_{k,\sigma}(\tau^{d\sigma}_{j}) c^{\dagger}_{k,\sigma}(\tau^{d^{\prime}\sigma}_{i})]}{Tr[e^{-\beta H_{c} }]}.
\end{eqnarray}

$w_{z}(\{\tau^{tot}\}_{n_{tot}} )$ comes from the $z$ component bosonic bath part \cite{pixley2011,pixley2013},
\begin{eqnarray}
w_{z}&=&\frac{Tr[e^{-\beta H_{B}^{z}} \prod_{i=1}^{n_{tot}}  e^{s_{i}\sum_{p} (g^{z}/\omega_{p}) ({\phi_{p}^{z}}^{\dagger}(\tau^{tot}_{i})-{\phi_{p}^{z}}(\tau^{tot}_{i} ))}  
  ]}
  {Tr[e^{-\beta H_{B}^{z}} ]}
  \nonumber \\
&=&\exp\left( -g^{2} \sum_{1<i<j<n_{tot}} s_{i}s_{j} \left(B(\tau_{i}-\tau_{j})-B(0)\right)   \right)
  \nonumber,
\end{eqnarray}
where $s_{i}=\pm s_{\sigma}$ or $\pm 1$ when the operator $O(\tau_{i}^{tot})$ at $\tau_{i}^{tot}$ corresponds to $d_{\sigma}^{\dagger} / d_{\sigma}$ or $S^{\pm}$, and
\begin{eqnarray}
B(\tau_{j}-\tau_{i})&=&
\sum_{p} \frac{ Tr[T_{\tau} e^{-\beta H_{B}^{z}} 
\phi^{z}_{p}(\tau_{i}) {\phi^{z}_{p}}^{\dagger}(\tau_{j}) ]}
{\omega_{p}^{2} Tr[e^{-\beta H_{B}^{z}} ]}
\nonumber \\
&+& (\tau_{i} \leftrightarrow \tau_{j}).
\end{eqnarray}

Finally,
$w_{p}(\{\tau^{s}\}_{m},\{\tau^{s^{\prime}}\}_{m})$ involves the bosonic bath in the transverse direction~\cite{otsuki},
forming a permanent,
\begin{eqnarray}
w_{p}&=&(g/\sqrt{2})^{2m} \left( \prod_{i=1}^{m }  \sum_{p_{i},p^{{\prime}}_{i}}  \right) Tr[e^{-\beta(H_{B}^{x}+H_{B}^{y})}T_{\tau}  \phi_{p_{m}}^{+}(\tau^{s^{\prime}}_{m})
\nonumber \\
&\times& \phi^{-}_{p^{\prime}_{m}}(\tau^{s}_{m})\cdots \phi^{+}_{p_{1}}(\tau^{s^{\prime}}_{1})\phi^{-}_{p^{\prime}_{1}}(\tau^{s}_{1}) ]/Tr[e^{-\beta(H_{B}^{x}+H_{B}^{y}) }]
\nonumber \\
&=&\sum_{p \in S_{m} } \prod_{i=1}^{m} P_{i,p(i)}.
\end{eqnarray}
The summation extends over $S_{m}$, representing all permutations of $1,2,\cdots,m$. 
The matrix element of $P$ is the following,
\begin{eqnarray}
P_{ij}&=&\frac{(g^{2}/2)\sum_{p}Tr[ e^{-\beta(H_{B}^{x}+H_{B}^{y})} T_{\tau} \phi^{-}_{p}(\tau^{s}_{j})  \phi^{+}_{p}(\tau^{s^{\prime}}_{i})]
}{Tr[e^{-\beta(H_{B}^{x}+H_{B}^{y})}]} \nonumber \\
&\equiv &(g^{2}/2) J(\tau_{j}^{s}-\tau_{i}^{s^{\prime}}).
\end{eqnarray}

Now the partition function can be interpreted as integrating a probability distribution function over some configuration space. Here, each configuration is specified by all sets of different $\{\tau^{\alpha} \}_{n}$ and a particular permutation $p \in S_{m}$, which is then sampled through a Metropolis algorithm with a probability proportional to $w_{d} \times w_{z} \times w_{c}^{\uparrow} \times w_{c}^{\downarrow} \times \prod_{i=1}^{m} P_{i,p(i)} $.

\begin{figure}[htb]
\captionsetup[subfigure]{labelformat=empty}
  \centering         
  \mbox{\includegraphics[width=0.95\columnwidth]{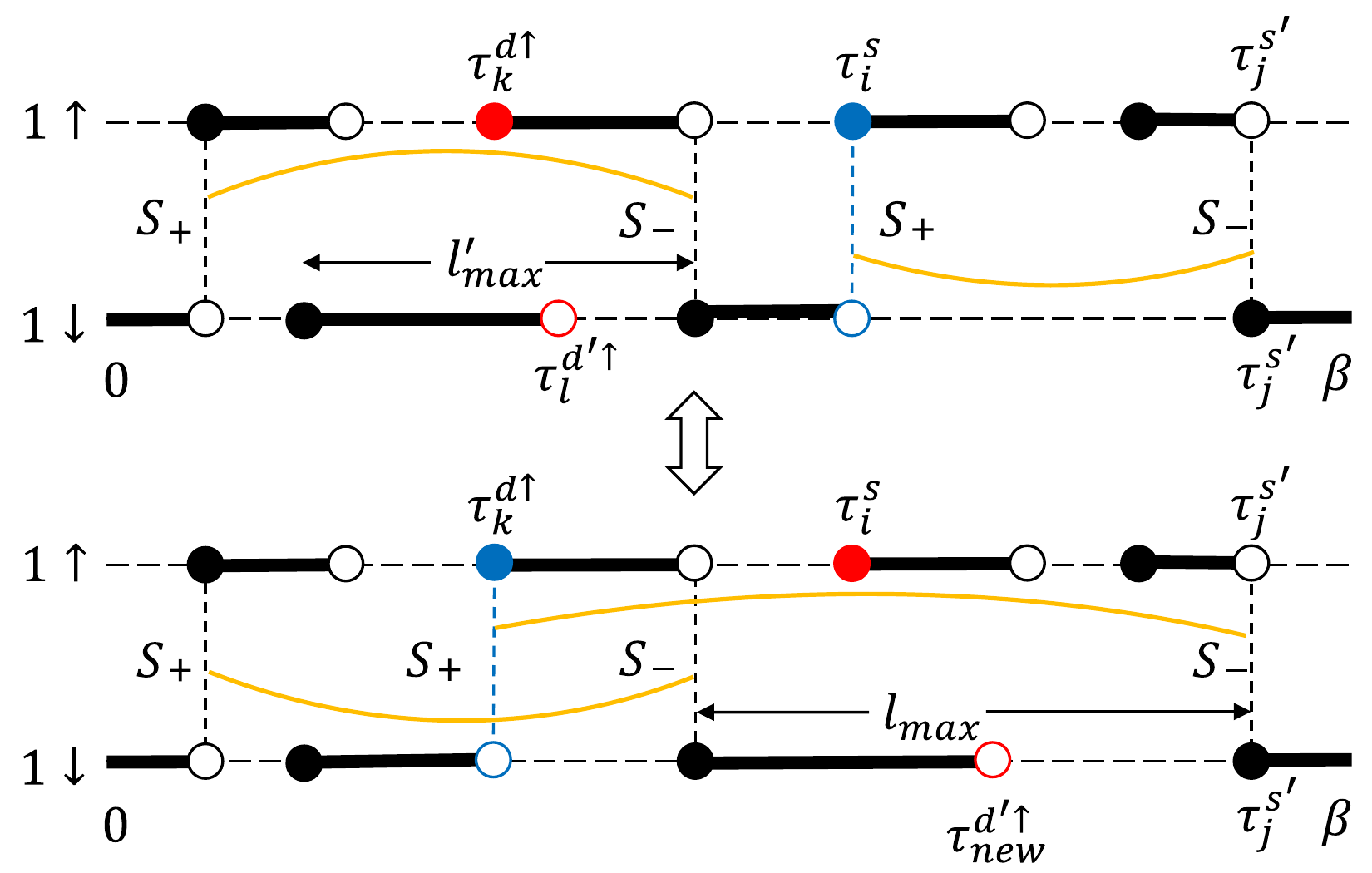}} 
\caption{\label{fig:update} Illustration of a swap update in a $m=2$, $n_{\uparrow}=2$, $n_{\downarrow}=1$ configuration. Filled/empty circles denote creation/annihilation operators along the imaginary time axis from $\tau=0$ to $\tau=\beta$. Vertical dashes lines label the composite $S_{+}$ and $S_{-}$ operators. Blue and red color denotes the affected $S_{+}$ and $d^{\dagger}_{\uparrow}$ and $d_{\downarrow}$ operator. Yellow lines specify the particular permutation in the permanent expansion.}
\end{figure}

We now describe the Monte Carlo updates. We inherit the updates from the Ising BFAM
\cite{pixley2011,pixley2013}, namely
 the insertion, removal and shift of  $d_{\sigma}^{\dagger} c_{k,\sigma}$/$c^{\dagger}_{k,\sigma} d_{\sigma}$ pair,
 and also  adopt the insertion/removal of $S^{+}\phi^{-}$/$S^{-}\phi^{+}$ and the sampling of the permutation 
 $S_{m}$ introduced in reference \cite{otsuki} (named updates (a)-(c) there).
 In addition we introduce 
 a swap update that swaps
  $S_{+} (S_{-}) $ with a pair of $ d_{\uparrow} ^{\dagger}$ and $d_{\downarrow} $ ( $ d_{\downarrow} ^{\dagger}$ and $d_{\uparrow} $  ). For example consider the $S_{+}$ case. We first randomly pick a pair of $S_{+}(\tau^{s}_{i})$, $S_{-}(\tau^{s^{\prime}}_{j})$ from the $m$ pairs of $S_{+}$ and $S_{-}$ that is connected by one of $J(\tau)$. Then we choose a $ d_{\uparrow}^{\dagger}(\tau^{d\uparrow}_{k})$ with a probability $ P_{k}=J (\tau^{d\uparrow}_{k}-\tau^{s^{\prime}}_{j}) / \left( \sum_{n=1,n_{\uparrow}} J(\tau^{d\uparrow}_{n} - \tau^{s^{\prime}}_{j} ) \right)$ from the $n_{\uparrow}$ of $d^{\dagger}_{\uparrow}$ operators. We then swap the position of  $S_{+}(\tau^{s}_{i})$ and $d_{\uparrow}^{\dagger}(\tau^{d\uparrow}_{k}) $. Finally, we find the $d_{\downarrow} (\tau^{d^{\prime}\downarrow}_{l}) $ that is closest to $ d_{\uparrow}^{\dagger}(\tau^{d\uparrow}_{k})$ before the swap, and move it to $d_{\downarrow} (\tau^{d^{\prime}\downarrow}_{new}) $. $\tau^{d^{\prime}\downarrow}_{new}$ is randomly selected within an interval of length $l_{max}$, which is the distance between two creation operators in the $\sigma=\downarrow$ orbital next to $S_{+}$ before the swap. The corresponding proposal probability is given by
\begin{equation}
P_{prop}= \frac{1}{l_{max} m } \times \frac{ J (\tau^{d\uparrow}_{k}-\tau^{s^{\prime}}_{j}) } 
{\sum_{n=1}^{n_{\uparrow}} J(\tau^{d}_{n} - \tau^{s^{\prime}}_{j} ) }.
\label{eq:P_prop}
\end{equation} 
Likewise we can find the proposal probability for the inverse update,
\begin{equation}
P_{prop}^{inv}=  \frac{1}{ l^{\prime}_{max} m} \times \frac{ J (\tau^{s}_{i}-\tau^{s^{\prime}}_{j} ) } 
{\sum_{\substack{ n=1, \\ n\neq k } }^{n_{\uparrow}} J(\tau^{d}_{n} - \tau^{s^{\prime}}_{j} ) + J(\tau^{s}_{i}- \tau^{s^{\prime}}_{j} ) }.
\label{eq:P_prop_inv}
\end{equation} 

The weight ratio between the proposed configuration and the current configuration is given by
\begin{eqnarray}
\frac{w_{new}}{w_{old}} &=& \frac{  w^{\uparrow}_{c}( \{ \tau^{d\uparrow} \}^{new}_{n_{\uparrow}} , \{ \tau^{d^{\prime}\uparrow} \}_{n_{\uparrow}}  ) w^{\downarrow}_{c}( \{ \tau^{d\downarrow} \}_{n_{\downarrow}} , \{ \tau^{d^{\prime}\downarrow} \}^{new}_{n_{\downarrow}}  )  }{  w^{\uparrow}_{c}( \{ \tau^{d\uparrow} \}_{n_{\uparrow}} , \{ \tau^{d^{\prime}\uparrow} \}_{n_{\uparrow}}  ) w^{\downarrow}_{c}( \{ \tau^{d\downarrow} \}_{n_{\downarrow}} , \{ \tau^{d^{\prime}\downarrow} \}_{n_{\downarrow}}  )   }
\nonumber \\
&\times & \frac{w_{d}( \{ \tau^{tot} \}^{new}_{n_{tot}} ) w_{z}( \{ \tau^{tot} \}^{new}_{n_{tot}} ) J (\tau^{d\uparrow}_{k}-\tau^{s^{\prime}}_{j}) }{w_{d}( \{ \tau^{tot} \}_{n_{tot}} )  w_{z}( \{ \tau^{tot} \}_{n_{tot}} ) J (\tau^{s}_{i} -\tau^{s^{\prime}}_{j}) }.
\label{eq:weight_ratio}
\end{eqnarray}
where $\{ \tau^{d\uparrow} \}^{new}_{n_{\uparrow}}$ is $\{ \tau^{d\uparrow} \}_{n_{\uparrow}}$ with $\tau_{k}^{d\uparrow}$ replaced by $\tau_{i}^{s}$, $\{ \tau^{d\downarrow} \}^{new}_{n_{\downarrow}}$ is $\{ \tau^{d\downarrow} \}_{n_{\downarrow}}$ with $\tau_{l}^{d^{\prime}\downarrow}$ replaced by $\tau_{lnew}^{d^{\prime}\downarrow}$, and $\{ \tau^{tot} \}^{new}_{n_{tot}}$ is $ \{ \tau^{tot} \}_{n_{tot}} $ with the above two substitutions, plus $\tau_{i}^{s}$ replaced by $\tau_{k}^{d\uparrow}$.

The detailed balance condition is satisfied by the adopting the acceptance ratio $max[R,1]$, with $R$ given by
\begin{equation}
R=\frac{w_{new}}{w_{old}} \times \frac{P_{prop}^{inv}}{P_{prop}}.
\end{equation}

The reason that we choose the proposal probability to be the form in equation~(\ref{eq:P_prop}) and equation~(\ref{eq:P_prop_inv}) is to cancel out the $ J (\tau^{d\uparrow}_{k}-\tau^{s^{\prime}}_{j})/J (\tau^{s}_{i} -\tau^{s^{\prime}}_{j})  $ factor in the weight ratio in equation~(\ref{eq:weight_ratio}), such that the acceptance ratio $R$ is of order 1. Otherwise if we select $ d^{\dagger}_{\uparrow} ( \tau_{k}^{d\uparrow})$ using a uniform distribution from $0$ to $\beta$, since $J(\tau) \sim 1/{\tau^{1+s}}$, on average $J(\tau^{d\uparrow}_{k}-\tau^{s^{\prime}}_{j}) \sim 1/ \beta^{s}$, while the average value of $ J (\tau^{s}_{i} -\tau^{s^{\prime}}_{j}) $ is $\beta$ independent, as a result $R$ will be suppressed by a factor of $1/\beta^{s}$. Similar ideas have been introduced in reference~\cite{steiner2015double}.

In practice we have tested that the swap update introduced here can replace the role of update (d) in reference~\cite{otsuki}, which breaks up one $S^{+}$ ($S^{-}$) into a pair of $d^{\dagger}_{\uparrow}$ and $d_{\downarrow}$ ( $d^{\dagger}_{\downarrow}$ and $d_{\uparrow}$ ) at two different time. Both of these updates are introducing shortcuts between configurations that are connected by a large number of other updates. But unlike update (d) whose acceptance ratio decreases with $\beta$ as a power-law, the swap update has an acceptance ratio that does not depend on $\beta$. This facilitates the task of reaching low enough temperatures and access the scaling 
regime. We have verified that our procedure preserves the SU(2) symmetry.

We now make a few remarks on how to evaluate $J(\tau)$ and $B(\tau)$ in the numerical calculation. This is important because in the current expansion scheme the weight contribution from the bosonic bath in the transverse direction $\phi^{\pm}$ and in the $z$ direction $\phi^{z}$ enters differently. Thereby the SU(2) symmetry of the model has to be recovered dynamically in the sampling process. In actual calculation we find that in order to maintain the SU(2) symmetry, it is crucial to evaluate $B(\tau)$ and $J(\tau)$ to sufficiently high accuracy.

Starting with the Fourier components of $J(\tau)$ in the matsubara frequency domain,
\begin{equation}
J(i\nu_{n})=\sum_{p} \frac{ 2 \omega_{p} }{{\omega_{p} }^{2}-\nu_{n}^{2}},
\end{equation}
where $\nu_{n}=2\pi n/\beta,\ n \in \mathbb{Z}$ is the matsubara frequencies. There are two ways to calculate $J(\tau)$. We can either perform the integration over the density of states first,
\begin{equation}
J(i\nu_{n} ) = \int_{0}^{\infty} \frac{2 \omega}{\omega^{2}-\nu_{n}^{2} }  \rho_{b}(\omega) d \omega,
\label{eq:mat_dos_int}
\end{equation}
followed by the matsubara summation,
\begin{equation}
J(\tau)=\frac{1}{\beta} J(i \nu_{n} =0 ) + \frac{2}{\beta} \sum_{\nu_{n}>0} J(i\nu_{n}) \cos (\nu_{n} \tau) .
\label{eq:mat_sum}
\end{equation}
Or we can first do the matsubara summation, then integrate over the density of states,
\begin{equation}
J (\tau) =\int_{0}^{\infty} \frac{ e^{(\beta-\tau)\omega} + e^{\tau \omega} }{ e^{\beta \omega} -1 } \rho_{b}(\omega) d \omega .
\label{eq:dos_int}
\end{equation}
In practice we find the summation in equation~(\ref{eq:mat_sum}) converges too slow when $\beta$ is large. So using equation~(\ref{eq:dos_int}) is recommended.

On the other hand, $J(\tau)$ is related to $B(\tau)$ by being its second derivative: $J(\tau)=d^{2} B(\tau) /d \tau^{2}$. $B(\tau)$ is most easily evaluated using the following formula,
\begin{eqnarray}
&& B(\tau)-B(0) \nonumber \\
&=& J(i\nu_{n}=0) \frac{\tau(\tau-\beta)}{2\beta} 
+ \sum_{n \neq 0} J (i\nu_{n}) \frac{1-\cos(\nu_{n}\tau)}{\beta \nu^{2}_{n}} \ \ \ .
\end{eqnarray}
Because of the extra $1/\nu_{n}^{2}$ factor here, the summation actually converges very quickly.

\subsection{Observables}
In this subsection we introduce all the quantities we will calculate using CT-QMC.

We start with the local magnetization,
\begin{equation}
\langle m_{\alpha} \rangle = \langle \frac{1}{\beta} \int_{0}^{\beta} { S} _ {\alpha} (\tau) d\tau \rangle , \ \alpha=x,y,z,
\end{equation}
which is related to most of the quantities we discussed below.

Because the sampling will preserve spin rotation symmetry, the actual measured $\langle m_{\alpha} \rangle $ is always 0. Instead we measure its root mean square,
\begin{equation}
\sigma_{\alpha}= \sqrt{ \langle m_{\alpha} ^{2} \rangle  },
\label{eq:sigma_def}
\end{equation}
which is also related to the static spin susceptibility 
$\chi^{spin}_{\alpha} (T) = \int_{0}^{\beta} \chi^{s}_{\alpha}(\tau)  d\tau=  \int_{0}^{\beta}  \langle T_{\tau} { S} _ {\alpha} (\tau)S _ {\alpha} \rangle  d\tau $
by,
\begin{equation}
\chi^{spin}_{\alpha}=\beta \sigma_{\alpha}^{2},
\label{eq:sigma_chi}
\end{equation}
where we have also defined the dynamical spin correlation function $\chi^{s}_{\alpha}(\tau)$. From $\chi^{s}_{\alpha}(\tau)$ we can also extract the spin correlation length $\xi_{\alpha}$ along the imaginary-time axis,
\begin{equation}
\xi_{\alpha}=\frac{1}{\nu_{1}} \sqrt{ \frac{\chi^{s}_{\alpha}(\nu_{0})}{ \chi^{s}_{\alpha}(\nu_{1}) } -1}.
\label{eq:xi_formula}
\end{equation}
Here $\chi^{s}_{\alpha}(\nu_{n})$ is the Fourier transform of $\chi^{s}_{\alpha}(\tau)$. This is in close analogy with extracting the
spatial
correlation length from the momentum dependence of the structure factor \cite{sandvik}. This can be understood by considering an ansatz $\chi^{s}_{\alpha}(\nu_{n})\propto (\nu_{n}^{2} + {E^{*}}^{2} )^{-x/2}$. At criticality, the crossover energy scale $E^{*}$ vanishes and $\chi^{s}_{\alpha}(\nu_{n})$ will diverge as $\chi^{s}_{\alpha}(\nu_{n})\propto \nu_{n}^{-x} $. Away from criticality $E^{*}$ is finite and will contribute a factor $ e^{-E^{*} \tau}$ to $\chi_{\alpha}^{s}(\tau)$ when transformed back to imaginary-time domain. Then using equation~(\ref{eq:xi_formula}) the crossover scale $E^{*}$ is inversely proportional to the correlation length $\xi_{\alpha} \propto 1/E^{*} $.

As we will always preserve spin SU(2) symmetry, in the following we will drop the subscript $\alpha$ labeling different spin components in any vector quantity.

We will also look at the Binder cumulant \cite{binder}, generalized to a n-components order parameter \cite{sandvik},
\begin{equation}
U_{2}=\frac{n+2}{2}\left( 1-\frac{n}{n+2} \frac{ \langle ( {\bf m} \cdot {\bf m} )^{2} \rangle }{ \langle {\bf m} \cdot {\bf m}  \rangle^{2} } \right),
\end{equation}
which is defined such that $U_{2}$ approaches $1$ in the ordered state and $0$ in the disordered state. Quantities like $\langle ( {\bf m} \cdot {\bf m} )^{2} \rangle$ will involve 4-point correlation functions of different components of $ S_{\alpha} $ which would require implementing worm type algorithm \cite{gunacker2015,gunacker2016}. 
In the presence of spin SU(2) symmetry, we can utilize the relation $\langle ( {\bf m} \cdot {\bf m} )^{2} \rangle=5 \langle m_{z}^{4} \rangle $ and $\langle {\bf m} \cdot {\bf m}  \rangle = 3\langle m_{z}^{2} \rangle$ to simplify the expression (for n=3),
\begin{equation}
U_{2}=\frac{5}{2} \left( 1- \frac{1}{3} \frac{ \langle m_{z}^{4} \rangle }{ \langle m_{z}^{2} \rangle^{2} }  \right).
\end{equation}

Another interesting quantity is the fidelity susceptibility $\chi^{\lambda}_{f}$. Suppose the Hamiltonian is composed of two parts $H=H_{\lambda=0}+\lambda H_{\lambda}$, with $\lambda$ being some tuning parameter. Then $\chi^{\lambda}_{f}$ is defined as \cite{albuquerque2010}
\begin{equation}
\chi^{\lambda}_{f}(T)=\int_{0}^{\beta/2} \left( \langle T_{\tau} H_{\lambda}(\tau) H_{\lambda} \rangle - \langle H_{\lambda} \rangle^{2} \right) \tau   d\tau.
\label{eq:fidel_original}
\end{equation}

At a second order quantum phase transition,
$\langle :H_{\lambda}(\tau) : : H_{\lambda} : \rangle \sim \left( {1}/{\tau} \right) ^{2 Dim[H_{\lambda}]}$.
Here $:H_{\lambda}:$ denotes normal ordering $:H_{\lambda}:=H_{\lambda}-\langle H_{\lambda} \rangle$ and $Dim[H_{\lambda}]$ denotes scaling dimension of $H_{\lambda}$. As we require $\int_{0}^{\beta} d\tau \lambda H_{\lambda}$ is scale invariant, we have $Dim[H_{\lambda}]=1-Dim[\lambda]$, so 
$\langle :H_{\lambda}(\tau) : : H_{\lambda} : \rangle \tau \sim \left( {1}/{\tau} \right) ^{1-2 Dim[\lambda]}$. We see that if $\lambda$ is relevant at the critical point, in which case it is usually identified as the correlation length exponent $\nu^{-1}$, $Dim[\lambda]=\nu^{-1}>0$, then $\chi^{\lambda}_{f}(T)$ will diverge,
\begin{equation}
\chi^{\lambda}_{f}(T) \propto \beta^{2/\nu}.
\end{equation}

Therefore $\chi^{\lambda}_{f}$ can be used to detect the location of a QPT, without knowing the actual order parameter. It turns out for hybridization expansion CT-QMC, if we choose $\lambda$ to be the hybridization strength $V$, then the corresponding fidelity susceptibility, which we denoted by $\chi^{V}_{f} $, can be calculate by a very simple formula \cite{lei.wang1,lei.wang2},
\begin{equation}
\chi^{V}_{f}=\frac{ \langle k_{L} k_{R} \rangle - \langle k_{L} \rangle \langle k_{R} \rangle  }{ 2 V^{2}  },
\label{eq:fidel}
\end{equation}
where we have considered dividing the imaginary-time axis into two pieces, and $k_{L}$ and $k_{R}$ are the number of $H_{\lambda}$ between $[0,\beta/2]$ and $[\beta/2,\beta]$ at each Monte Carlo step, respectively.

\section{Quantum Phase Transitions and Phase Diagram}
\label{numerical}

We now present the CT-QMC results. We describe the details of our analysis in the representative cases
of $s=0.6$ in section~\ref{numerical.A} and $s=0.2$ in section~\ref{numerical.B}. We then consider the dependence
on $s$ in the range $0<s<1$ appropriate for sub-ohmic bosonic bath in section~\ref{numerical.C}.

\subsection{s=0.6}
\label{numerical.A}
We start by presenting our analysis at $s=0.6$, which belongs to the case of RG flow specified in figure~\ref{fig:rg} (a).
Alongside with C$^{\prime}$ that controls the transition from local moment phase to Kondo phase, 
due to the appearance of a stable fixed point L representing the critical phase, 
we have two additional unstable fixed points C and LC, 
each describing the transition from critical phase to Kondo phase and critical phase to local moment phase, 
respectively. In the following, we will present numerical evidence for each of the three QCPs.

\subsubsection{Critical phase-Kondo transition}
\label{L-Kondo}
\begin{figure}[!htb]
\captionsetup[subfigure]{labelformat=empty}
  \centering
  \mbox{\includegraphics[width=0.85\columnwidth]{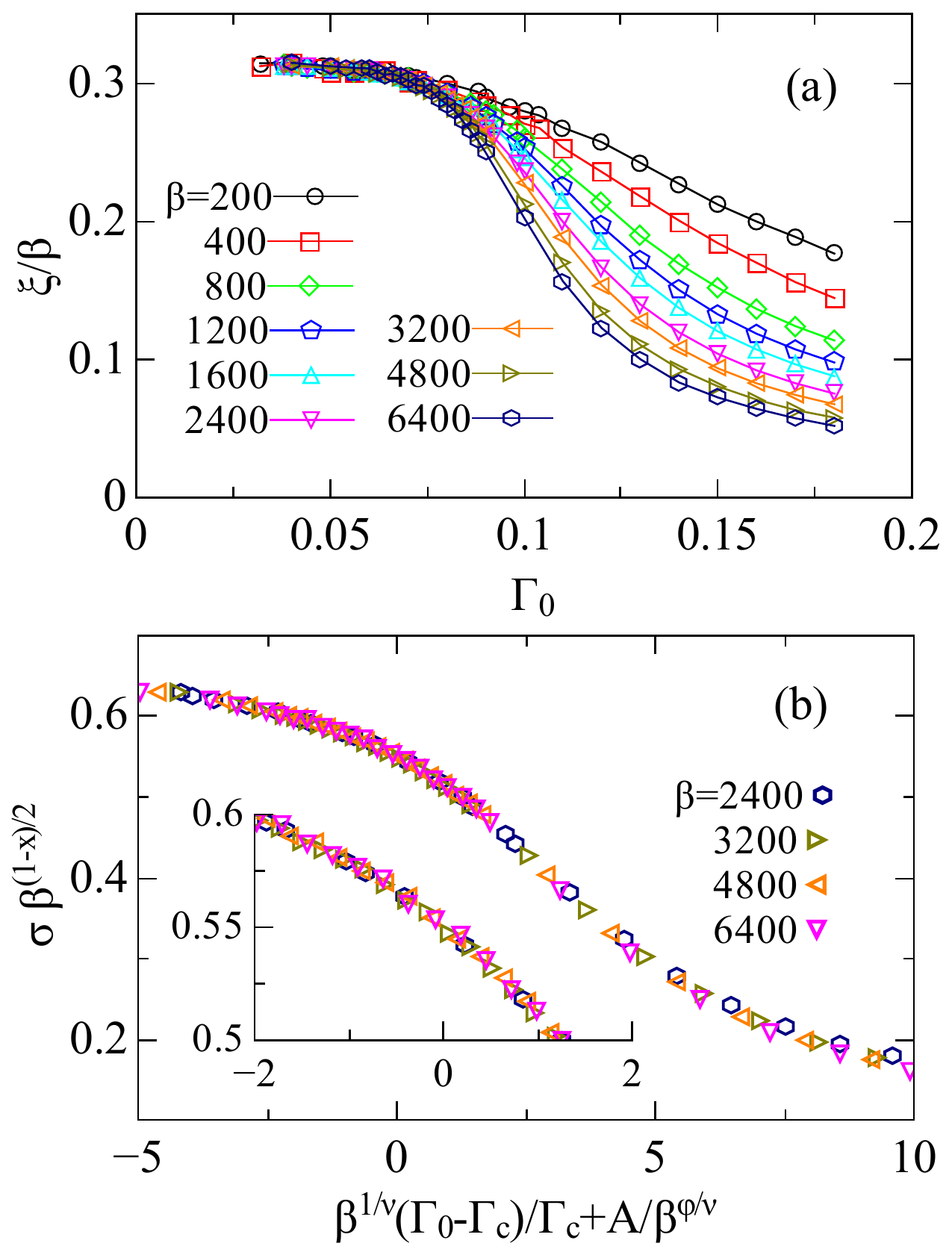}}    
\caption{(a) Reduced correlation length vs. $\Gamma_{0}$ from $\beta=200$ to $\beta=6400$ at $g=0.5$, $s=0.6$. $\xi/\beta$ remains constant in the critical phase while decreases with increasing $\beta$ in the Kondo phase. (b) Rescaled magnetization using equation (\ref{eq:fss_sigma}), with $x=0.63(2)$, $\Gamma_{c}=0.08(1)$ and $ \nu^{-1}=0.26(4)$. Inset: blow up view near $\Gamma_{0}=\Gamma_{c}$. }
\label{fig:5}
\end{figure}

First we stay at $g=0.5$, and gradually increase $\Gamma_{0}$.
In figure~\ref{fig:5}(a) we plot $\xi/\beta$ versus $\Gamma_{0}$ from $\beta=200$ all the way to $\beta=6400$. For $\Gamma_{0} \lesssim 0.08$, we find $\xi/\beta$ is almost independent of $\beta$ (system size), suggesting the system being scale invariant for a range of $\Gamma_{0}$. This is the signature of the critical phase. At larger $\Gamma_{0}$, $\xi$ grows slower than the system size $\beta$, signifying short time correlation between the impurity spin as the impurity is Kondo screened. At some critical value of $\Gamma_{0}$ we expect a quantum phase transition separating the two phases. But the exact location is hard to pin-point, as we do not see any crossing in $\xi/\beta$. In section \ref{L'-Kondo} we will show that $\xi/\beta$ does have a crossing at the local moment to Kondo QCP.

One observable we can utilize is the root mean square magnetization $\sigma$ defined in equation~(\ref{eq:sigma_def}). We expect a scaling form as follows should hold,
\begin{equation}
\sigma(\Gamma_{0},\beta) = \beta^{-(1-x)/2} \tilde{\sigma} \left(\beta^{1/\nu}(\Gamma_{0}-\Gamma_{c})/\Gamma_{c} + A/\beta^{\phi/\nu} \right).
\label{eq:fss_sigma}
\end{equation}
where $\tilde{\sigma}$ is the universal function and $A/ \beta^{\phi/\nu}$ is the sub-leading terms.

In the universal function $ \tilde{\sigma}$ the dependence of the tuning parameter only comes in through the combination of $\beta^{1/\nu} (\Gamma_{0}-\Gamma_{c})$ (ignoring sub-leading corrections). This can be justified from RG or understood phenomenologically bases on the consideration that at QCP the system only depend on the ratio $\beta/\xi$ and $\xi$ diverges with $\xi \propto |\Gamma_{0}-\Gamma_{c}|^{-\nu}$. One subtlety here is that the correlation length diverges in the entire critical phase. So one could question whether such a scaling form still apply in the region of $\Gamma_{0}<\Gamma_{c}$.  The prefactor $\beta^{-(1-x)/2}$ comes from equation~(\ref{eq:sigma_chi}) and that at the QCP we expect $\chi^{spin} \propto \beta^{x} $ with the exact relation $x=s$ based on $\epsilon$-expansion RG result. Here instead of imposing this relation we allow $x$ to adjust freely. As shown in figure~\ref{fig:5}(b), the quality of the scaling collapse suggests that equation~(\ref{eq:fss_sigma}) is the correct scaling hypothesis. In addition the correspondingly determined $\Gamma_{c}=0.08(1)$ and $\nu^{-1}=0.26(3)$ are consistent with what we obtained from $\chi^{V}_{f}$. We also find $x=0.63(4)$, consistent with the prediction $x=s$. 

From $\epsilon$-expansion calculation to second order \cite{zhu2002,zarand2002}, we obtain $\nu^{-1}= \epsilon/2 + \epsilon^{2} /6 \simeq 0.23$, in reasonably good agreement with the numerical value.
 
\begin{figure}[!htb]
\captionsetup[subfigure]{labelformat=empty}
  \centering
  \mbox{\includegraphics[width=0.8\columnwidth]{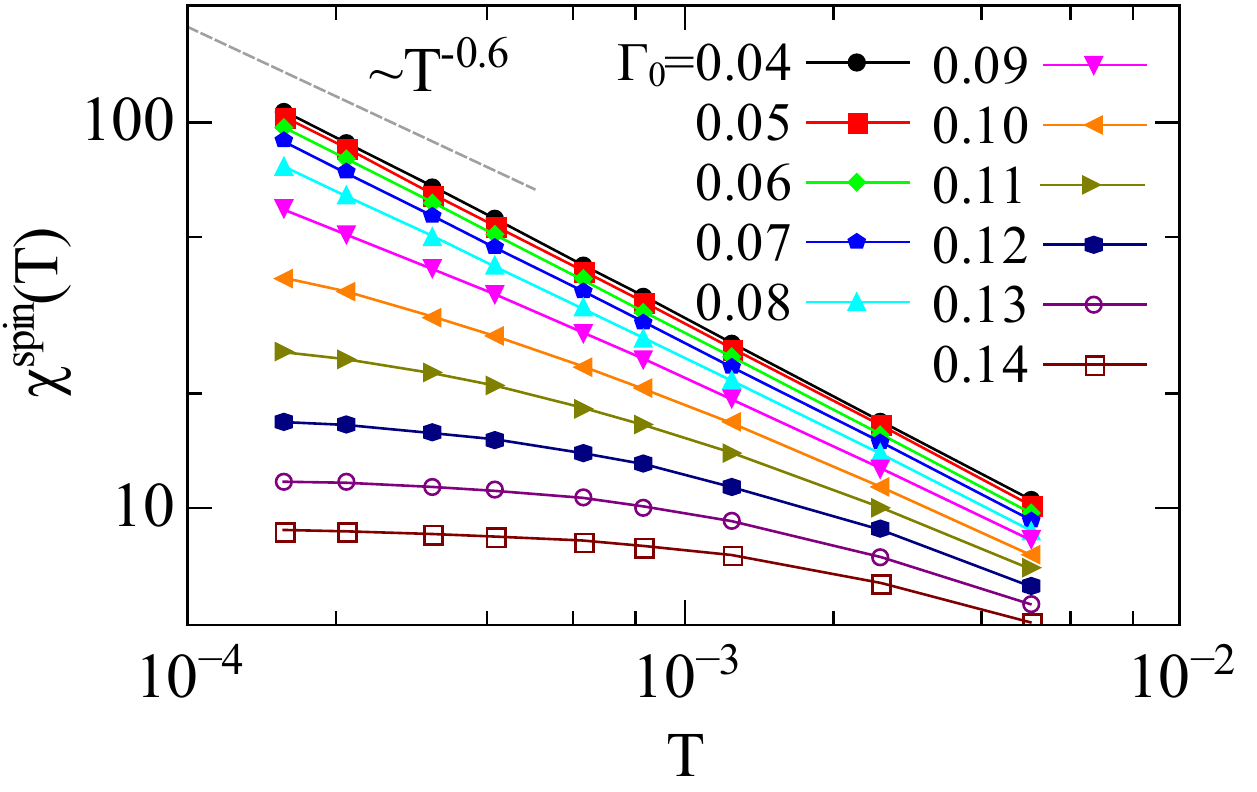}}    
\caption{\label{fig:7} Temperature dependence of spin susceptibility across the critical phase to Kondo QCP at $g=0.5$, $s=0.6$. Dashed line shows the $T^{-s}$ behavior expected in the critical phase $( \Gamma_{0} \lesssim 0.08  )$ as well as at the QCP $( \Gamma_{0}\simeq 0.08 )$.}
\end{figure}

Unlike the $\chi^{spin}(T) \sim  1/T$ local moment behavior in the $s=0.2$ case previously found in reference \cite{otsuki}, here the temperature dependence of the spin susceptibility obeys a nontrivial power-law, as shown in figure~\ref{fig:7}. We find $x=0.66,0.67,0.66,0.65$ for $\Gamma_{0}=0.04,0.05,0.06,0.07$ respectively. We interpret this as all the $\Gamma_{0} < \Gamma_{c}$ points under RG will flow towards the critical phase fixed point L with $\chi^{spin}(T) \sim A_{1}/T^{s}$. Notice that according to $\epsilon$-expansion the leading irrelevant operator has a very small scaling dimension $y_{i}=-\epsilon/2+O(\epsilon^{2})$, so the deviation from the exact relation $x=s$ is most likely due to corrections to scaling. At $\Gamma_{0}=\Gamma_{c}$, we have $x=0.61$. This is also consistent with the predicted critical behavior $\chi^{spin}(T) \sim A_{2}/T^{s}$ at fixed point C from $\epsilon$-expansion RG.

\subsubsection{Critical phase-local moment transition}
So far we have considered the regime accessible by the $\epsilon$-expansion of the SU(2) model, 
namely when both the fermionic and bosonic couplings are small.
Unlike the Coulomb-gas expansion of the Ising case~\cite{si1996kosterlitz,smith1999,zhu2002,zarand2002},
the $\epsilon$-expansion here does not reach the regime of large $g$.
In order to simplify the calculation we set $\Gamma_{0}=0$ in this section. 
We have also performed calculation at small but
nonzero $\Gamma_{0}$ and the conclusion remains the same.

\begin{figure}[!htb]
\captionsetup[subfigure]{labelformat=empty}
  \centering
   \mbox{\includegraphics[width=0.85\columnwidth]{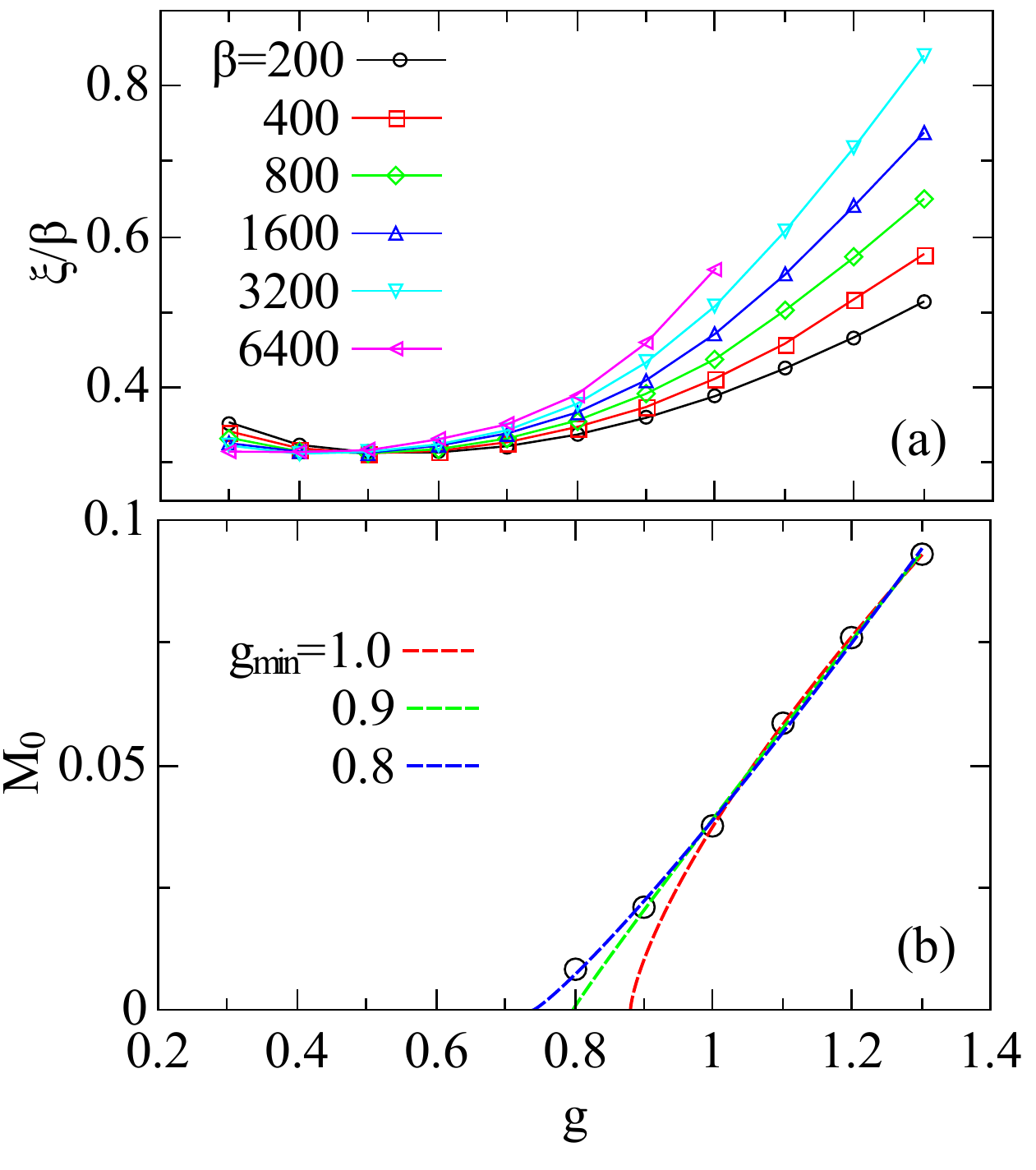}}                            
\caption{(a) Reduced correlation length as a function of $g$. The distinct behavior at small and large $g$ each corresponds to critical phase and local moment phase. (b) Effective Cuire constant extracted using equation (\ref{eq:chi_spin}) as a function of $g$. Dashed lines are power law fits according to $M_{0} \propto (g-g_{c})^{\beta_{1}}$ up to $ g \geq g_{min}$ with three different choice of $g_{min}$. }
\label{fig:8}
\end{figure}

First let us look at the behavior of the correlation length as a function of $g$, plotted in figure~\ref{fig:8}. The low temperature behavior of $\xi/\beta$ for $g\lesssim 0.5$ resembles the critical phase behavior in figure~\ref{fig:5}(a), both converging to a value around $0.3$. For $g\gtrsim 0.8$, on the other hand, $\xi/\beta$ rises as temperature decreases, which suggests local moment phase behavior. 

A more quantitative way of studying the transition between these two phases is by looking at the temperature dependence of the mean square magnetization $\sigma^{2}$. 
Following reference~\cite{otsuki}, the low temperature behavior of $\chi^{spin}(T)$ can be described by the following ansatz,
\begin{equation}
\chi^{spin}(T)={M_{0}}/{T} + {1}/{T^{x} T_{B}^{1-x}}.
\label{eq:chi_spin}
\end{equation}
Here $M_{0}$ is the Curie constant, $T_{B}$ the crossover temperature scale above which the critical fluctuation part $T^{-x}$ will dominate. This together with equation (\ref{eq:sigma_chi}) leads to
\begin{equation}
\sigma^{2}(T)=M_{0} + \left( {T}/{T_{B}} \right)^{1-x}.
\label{eq:sigma_vs_T}
\end{equation}

\begin{figure}[!htb]
\captionsetup[subfigure]{labelformat=empty}
  \centering
  \mbox{\includegraphics[width=0.9\columnwidth]{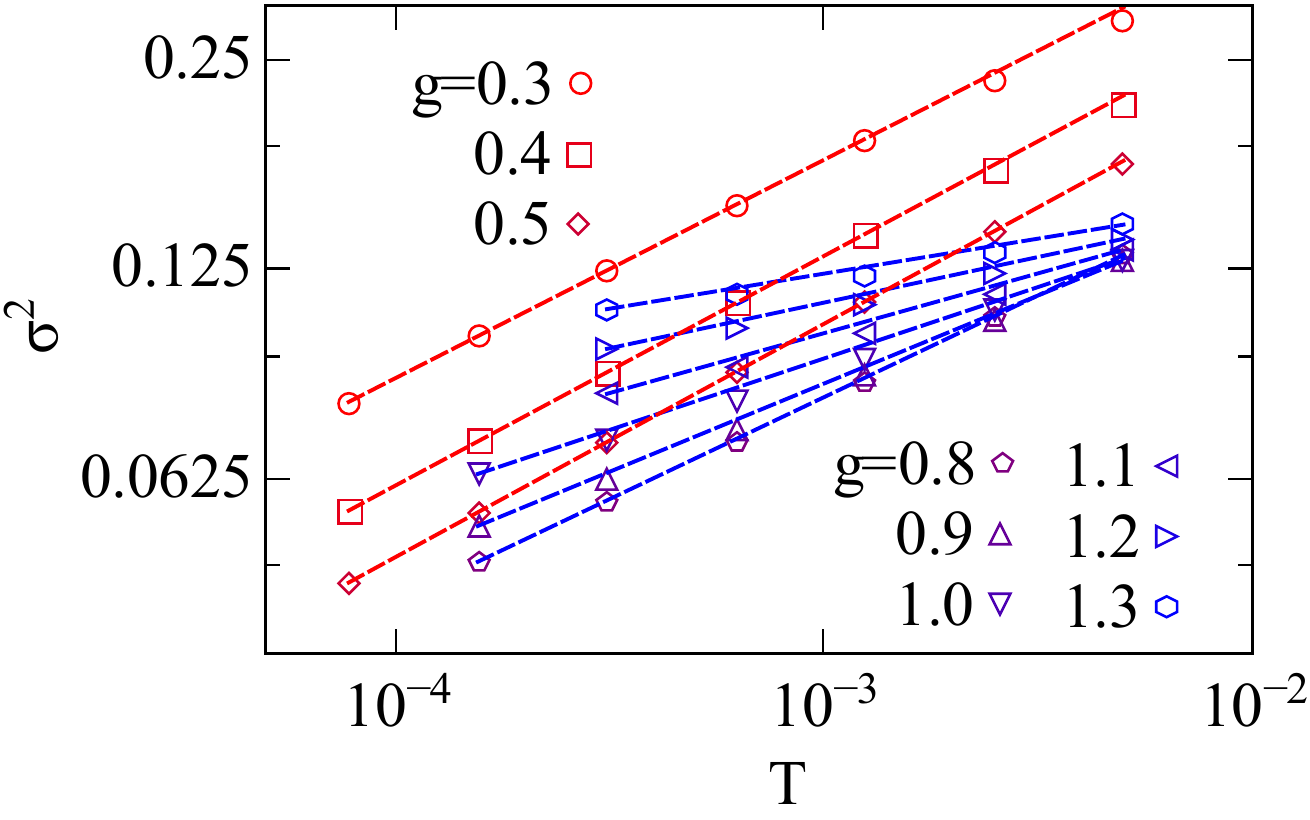}}    
\caption{Temperature dependence of mean square magnetization across the critical phase-local moment transition. Red (Blue) lines are fits according to equation~(\ref{eq:sigma_vs_T}) with zero (finite) curie constant $M_{0}$, which is expected in the critical (local moment) phase.}
\label{fig:9}
\end{figure}

Our result for $\sigma^{2}(T)$ is plotted in figure~\ref{fig:9}. 
For $g\leq 0.5$, the data can be described by equation~(\ref{eq:sigma_vs_T}) with $M_{0}=0$ and $x=0.68,0.67,0.66$ for $g=0.3,0.4,0.5$. This is the critical phase and the exponent is very close to what we obtained at Sec.\ref{L-Kondo}. For $g \geq 0.8$, fitting $\sigma^{2}(T)$ using the same equation gives a finite $M_{0}$. This indicates we are entering the local moment phase. While we have obtained $x=0.60$ for $g>1$, we have $x=0.67,0.65,0.64$ for $g=0.8, 0.9, 1.0$, reflecting corrections to scaling not captured by equation~(\ref{eq:sigma_vs_T}).

The extracted $M_{0}$ is plotted in figure~\ref{fig:8}(b). Close to the transition point at $g=g_{c}$, we expect $M_{0}$ will vanish as $M_{0} \propto (g-g_{c})^{\beta_{1}}$. We attempt to use this relation to find the value of $g_{c}$ by fitting over the $M_{0}$ versus $g$ data. Bearing in mind that for $0.8\leq g \leq 1$ the value of $x$ obtained from equation~(\ref{eq:sigma_vs_T}) is larger than $s$, it is likely that we will be overestimating $M_{0}$ in this region, so we only use $M_{0}$ down to $g\geq g_{min}$, and vary $g_{min}$ from $0.8$ to $1$. Depending on the cutoff $g_{min}$, the obtained $g_{c}$ lands within the range $g_{c} \in [0.74,0.88]$. Notice that the fitting with different $g_{min}$ all describe the $g \geq 1$ part of the data quite well. We thus take our final estimate of $g_{c}$ to be $g_{c} = 0.8 \pm 0.1$.

\subsubsection{Local moment-Kondo transition }
\label{L'-Kondo}
\begin{figure}[!htb]
\captionsetup[subfigure]{labelformat=empty}
  \centering
  \mbox{\includegraphics[width=0.9\columnwidth]{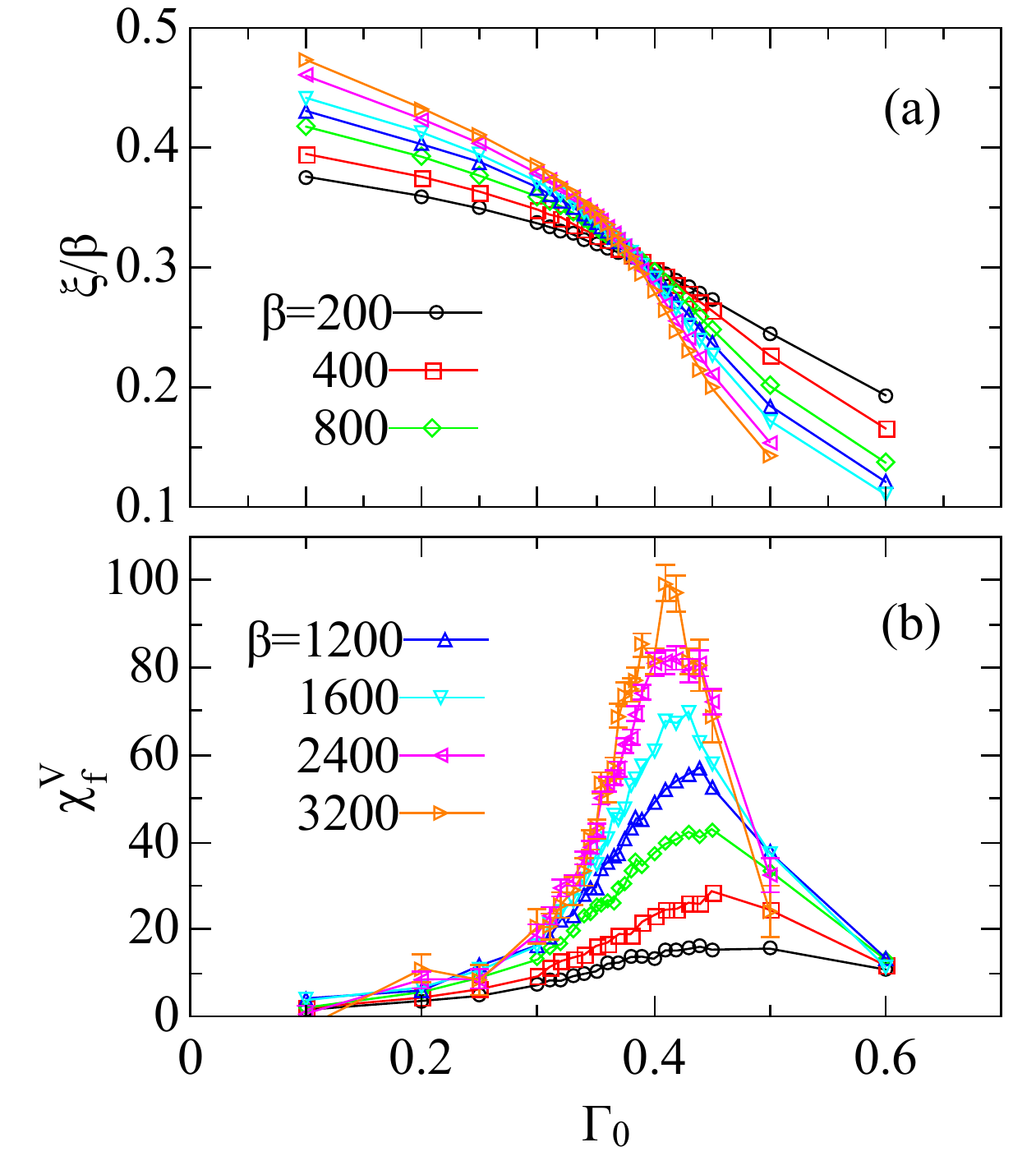}}    
\caption{Reduced correlation length $\xi/\beta$ (a) and fidelity susceptibility $\chi_{f}^{V}$ (b) vs. $\Gamma_{0}$ across the local moment-Kondo transition from $\beta=200$ to $\beta=6400$ at $g=1$, $s=0.6$. Near the QCP $\xi/\beta$ exhibits crossing and $\chi_{f}^{V}$ shows up a peak.}
\label{fig:s_6_g_1_1}
\end{figure}

Now that we have established that the system resides in the local moment phase for $g>g_{c} \simeq 0.8$ at $\Gamma_{0}=0$, we consider a path to the Kondo screened phase by turning on the hybridization while fixing $g=1$. As expected, we observe a crossing in $\xi/\beta$, and a divergence in $\chi_{f}^{V}$, both around $\Gamma_{0}=0.4$ (cf. figure~\ref{fig:s_6_g_1_1}). 

Similar to what we have done for $\sigma(\Gamma_{0},\beta)$ in equation~(\ref{eq:fss_sigma}), we consider the following finite size scaling hypothesis for $\xi$ and $\chi_{f}^{V}$,
\begin{eqnarray}
\xi(\Gamma_{0},\beta)&=&\beta \tilde{\xi}\left(\beta^{1/\nu}(\Gamma_{0}-\Gamma_{c})/\Gamma_{c}+A/\beta^{\phi/\nu} \right) \label{eq:2}, \\
\chi^{V}_{f}(\Gamma_{0},\beta) &=& \beta^{2/\nu}  \tilde{\chi} \left(\beta^{1/\nu}(\Gamma_{0}-\Gamma_{c})/\Gamma_{c}+A/\beta^{\phi/\nu} \right).
\label{eq:chif}.
\end{eqnarray}

\begin{figure}[!htb]
\captionsetup[subfigure]{labelformat=empty}
  \centering
  \mbox{\includegraphics[width=0.9\columnwidth]{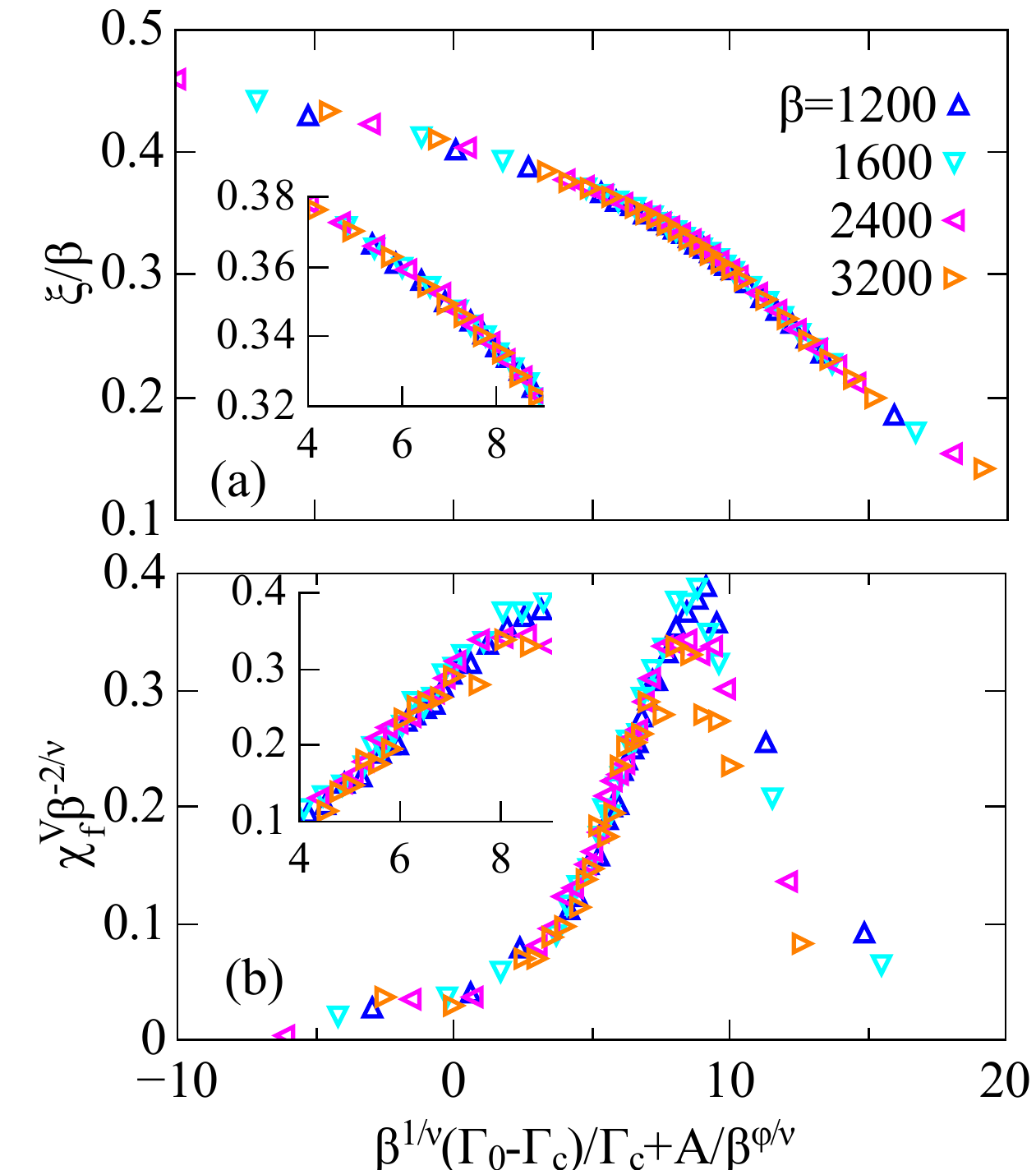}}    
\caption{Finite size scaling of correlation length (a) and fidelity susceptibility (b) for the local moment to Kondo transition. Inset shows blow up view of data obtained near $\Gamma_{0}=\Gamma_{c}$. Note that due to incorporating the sub-leading term in the scaling ansatz, it is no longer centered around $0$. }
\label{fig:s_6_g_1_2} 
\end{figure}

As seen in figure~\ref{fig:s_6_g_1_2}, close to the critical point the data fall nicely under a single universal curve. We obtain $\Gamma_{c}=0.35(2), \nu^{-1}=0.39(6)$ from $\xi$ and $\Gamma_{c}=0.34(2), \nu^{-1}=0.37(5)$ from $\chi_{f}^{V}$. Our final estimated value are $\Gamma_{c}=0.35(2)$ and $\nu^{-1}=0.38(5)$. The value of $\nu^{-1}$ obtained here for critical point C$^{\prime}$ is in sharp contrast with that for critical point C with $\nu=0.25(4)$. This further established that C and C$^{\prime}$ are two distinct critical points.

We now turn to the critical behavior of spin susceptibility. We plot $\chi^{spin}$ vs. $T$ at different $\Gamma_{0}$. At the critical coupling $\Gamma_{0}=\Gamma_{c} \simeq 0.35$, $\chi^{spin}(T)$ can be fitted with a power law $\chi^{spin}(T) \propto T^{-x} $ with $x=0.65$. Inside the local moment phase at $\Gamma_{0}=0.1$, it can be described by equation~(\ref{eq:chi_spin}) with a finite $M_{0}=0.10$ for the $M_{0}/T$ term and a sub-leading $1/T^{x}T_{B}^{1-x}$ term with $x=0.62$. These are consistent with the critical spin fluctuations being dominated by a $T^{-s}$ behavior. Thus we think the local spin susceptibility at C$^{\prime}$ should also diverge as $\chi^{spin} \sim 1/T^{s}$.

\begin{figure}[!htb]
\captionsetup[subfigure]{labelformat=empty}
  \centering
  \mbox{\includegraphics[width=0.9\columnwidth]{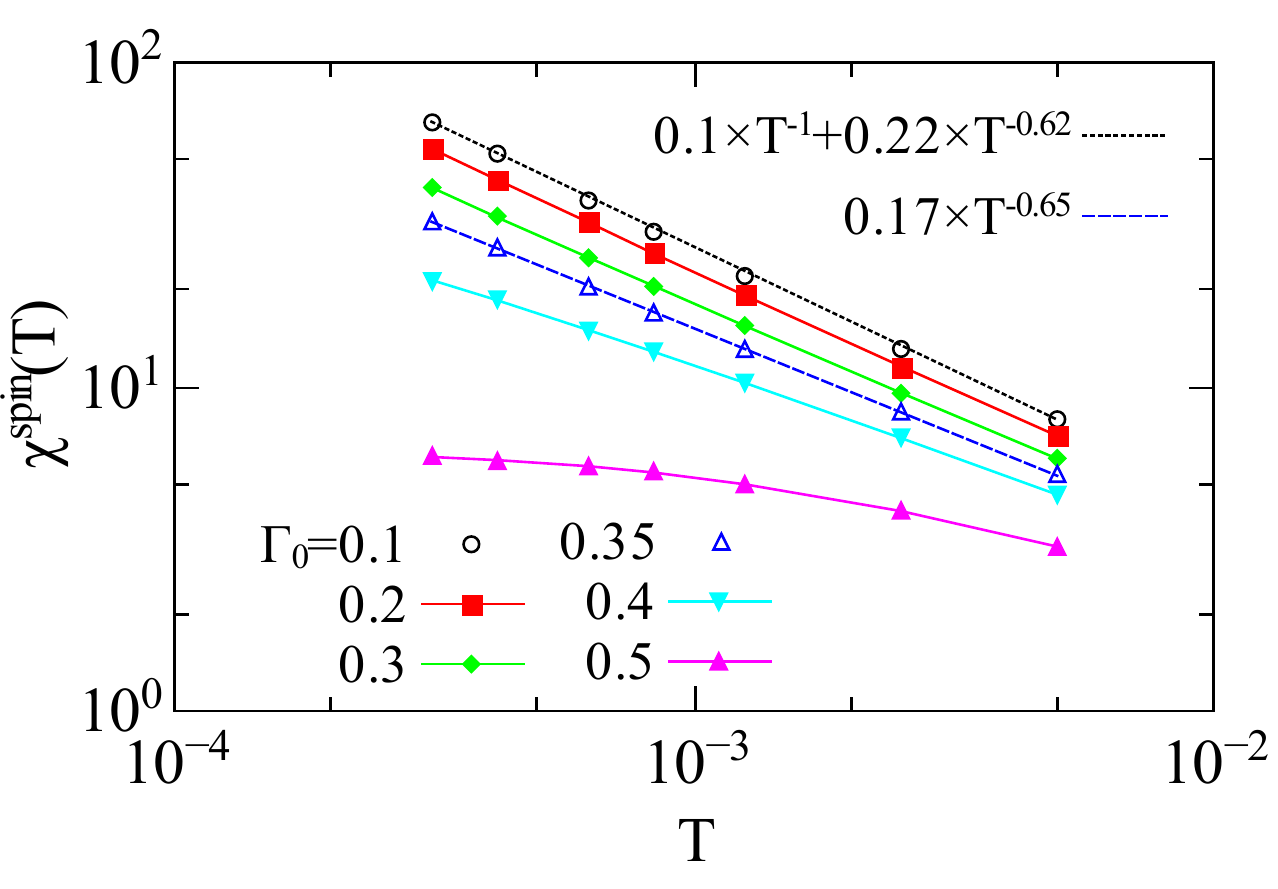}}    
\caption{Temperature dependence of spin susceptibility across the local moment to Kondo transition at $g=1, s=0.6$. Black dotted line is fit from data at $\Gamma_{0}=0.1$ in the local moment phase using equation~(\ref{eq:chi_spin}) with a finite Curie part $M_{0}=0.1$. Blue dashed line is fit from data at $\Gamma_{0}=\Gamma_{c}$ with power-law behavior $\chi^{spin}(T) \sim T^{-0.65} $.}
\label{fig:s_6_g_1_3} 
\end{figure}

\subsection{s=0.2}
\label{numerical.B}
We now turn to the $s=0.2$ case. 
This is also the case investigated in reference~\cite{otsuki} at the $U=\infty$ limit. We will fix $g=0.5$ and gradually increase $\Gamma_{0}$ to find the QCP from local moment phase to Kondo screened phase.

\begin{figure}[htb]
\captionsetup[subfigure]{labelformat=empty}
  \centering
  \mbox{\includegraphics[width=0.8\columnwidth]{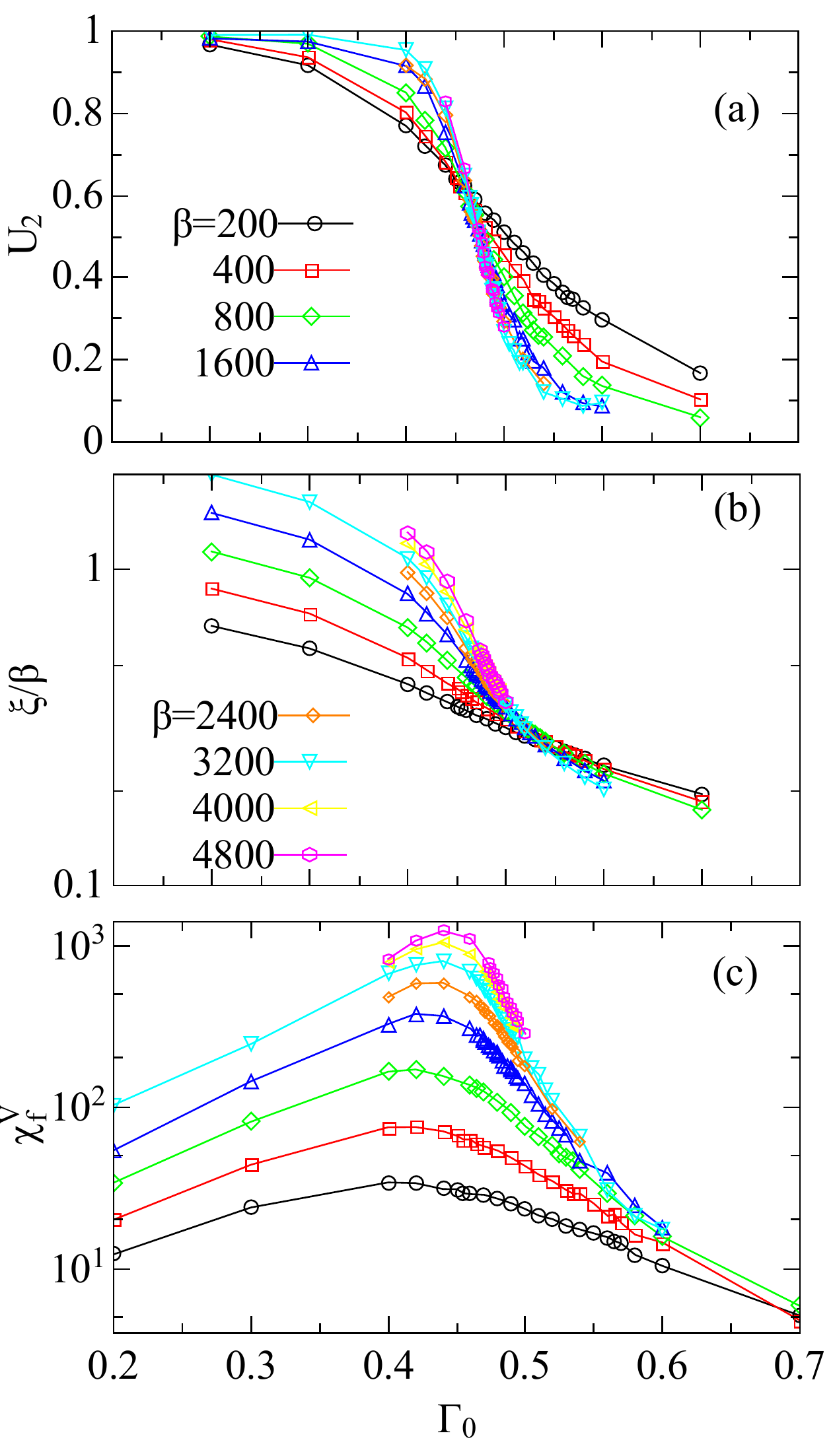}}     
     \\[0ex]                      
\caption{Various quantities vs hybridization strength $\Gamma_{0}$ across the local moment to Kondo QCP, including (a) binder cumulant (b) reduced correlation length and (c) fidelity susceptibility, from $\beta=200$ to $\beta=3200$ at $s=0.2$, $g=0.5$. Near the QCP $U_{2}$ and $\xi/\beta$ exhibits crossing while $\chi_{f}^{V}$ shows up a peak.}
\label{fig:1} 
\end{figure}

\begin{figure}[htb]
\captionsetup[subfigure]{labelformat=empty}
  \centering                        
    \mbox{\includegraphics[width=0.8\columnwidth]{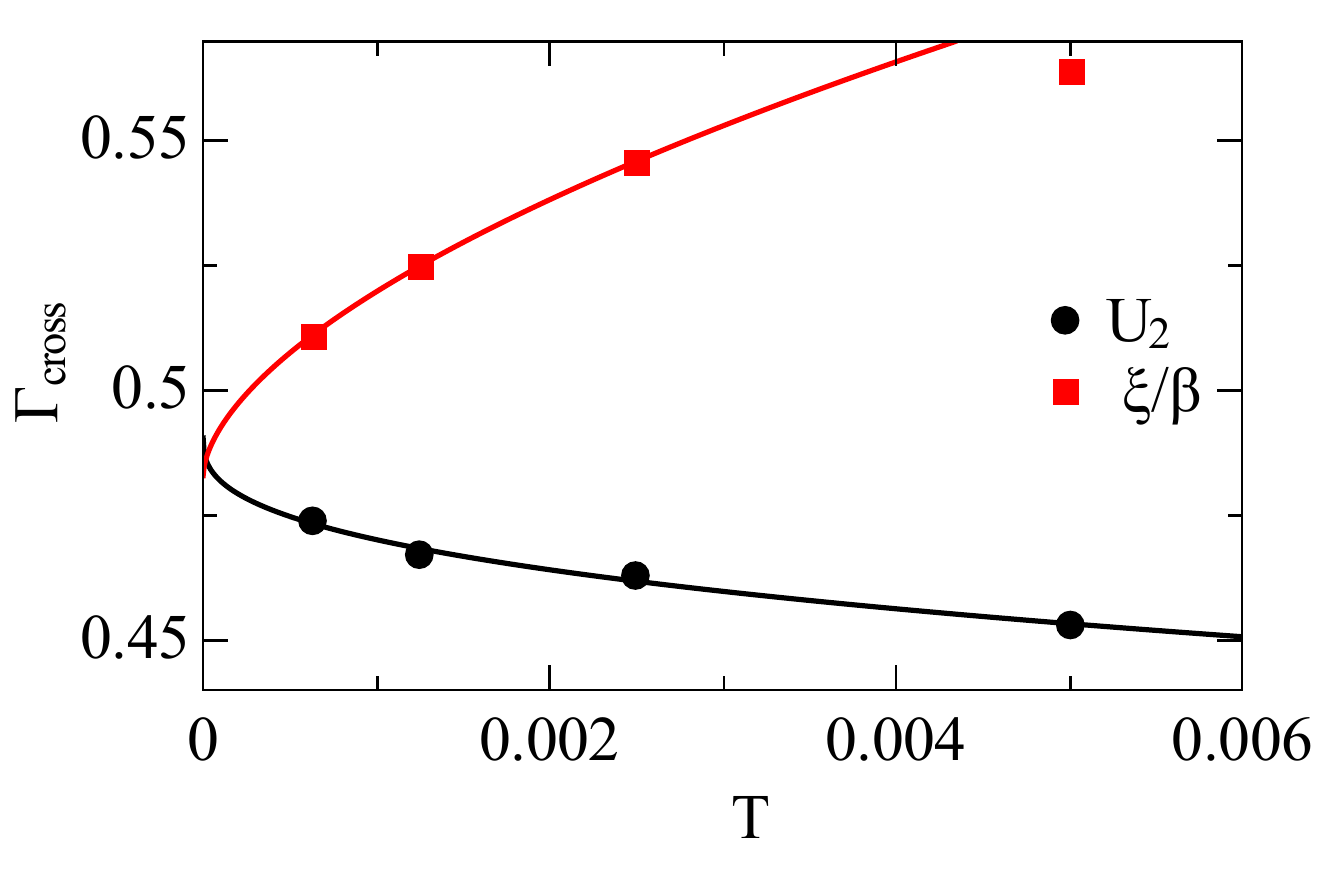}} 
\caption{Evolution of the crossing points in $U_{2}$ and $\xi/\beta$ as temperature is lowered. Data are extracted from fig. \ref{fig:1}(a)(b). Curves are fits to $\Gamma_{cross}=\Gamma_{c}+ aT^{b}$, showing that crossing points are converging to a common value.}
\label{fig:2}
\end{figure}

We first plot the dependence on $\Gamma_{0}$ of the Binder cumulant $U_{2}$ and the reduced correlation length $\xi/\beta$ in figure~\ref{fig:1}(a) and figure~\ref{fig:1}(b), where we have identified crossing points for both quantities. 
This suggests a transition from a local moment phase at small $\Gamma_{0}$ to a Kondo screened phase at large $\Gamma_{0}$. 
The crossing points have a sizable drift as we lower the temperature, which can be seen more clearly by plotting the crossing points between curves at $\beta$ and $2\beta$ in figure~\ref{fig:2}. 
We see that the crossing points obtained from $U_{2}$ and $\xi/\beta$ are approaching to the same critical value $\Gamma_{c}$ in the $T=0$ limit from opposite directions. By extrapolating the crossing points $\Gamma_{cross}$ to $T=0$ using a simple power-law relation $\Gamma_{cross}=\Gamma_{c}+aT^{b}$, we find $\Gamma_{c}=0.48(1)$.

We can then repeat the analysis done in Sec.\ref{L-Kondo} for the same type of transition at $s=0.2$ by considering scaling collapse of the form in equation~(\ref{eq:2}) for the correlation length $\xi$ and similarly for the Binder cumulant $U_{2}$, 
\begin{eqnarray}
U_{2}(\Gamma_{0},\beta)= \tilde{U}_{2} \left(\beta^{1/\nu}(\Gamma_{0}-\Gamma_{c})/\Gamma_{c}+A/\beta^{\phi/\nu} \right) \label{eq:1}
\end{eqnarray}
where the presence of the sub-leading term $A/\beta^{\phi/\nu}$ can take into account the finite temperature shift of the crossing point.

\begin{figure}[!htb]
\captionsetup[subfigure]{labelformat=empty}
  \centering
     \mbox{\includegraphics[width=0.9\columnwidth]{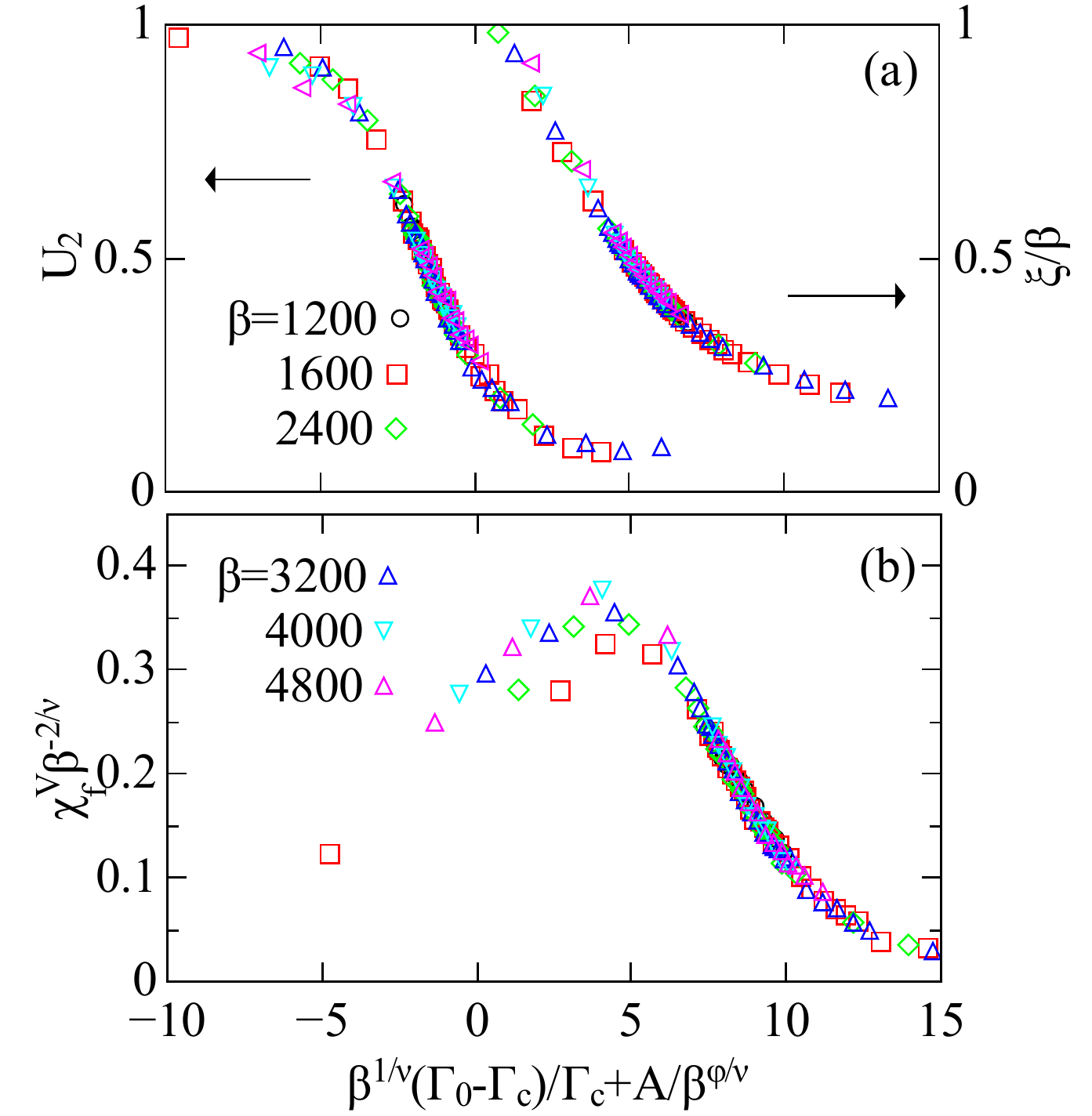}}                              
\caption{Finite size scaling analysis for critical point C$^{\prime}$ at s=0.2 based on data in fig.\ref{fig:3}. Upper panel: scaling collapse of Binder cumulant with $\Gamma_{c}=0.49(1)$, $\nu^{-1}=0.42(3)$ and correlation length with $\Gamma_{c}=0.48(1)$, $\nu^{-1}=0.43(3)$. Lower panel: scaling collapse of fidelity susceptibility with $\Gamma_{c}=0.46(2)$, $\nu^{-1}=0.48(3)$.}
\label{fig:3}
\end{figure}

It turns out these ansatzes describe the data very well. The collapsed data using equation~(\ref{eq:1}) and equation~(\ref{eq:2}) is plotted in figure~\ref{fig:3}(a), and they give consistent estimates for the value of the critical coupling $\Gamma_{c}$ and correlation length exponent $\nu$. We obtain $\Gamma_{c}=0.49(1)$, $\nu^{-1}=0.42(3)$ from $U_{2}$ and $\Gamma_{c}=0.48(1)$, $\nu^{-1}=0.43(3)$ from $\xi$. 

We further test the applicability of the fidelity susceptibility in this case, which serves as another independent tool to detect the QCP. As shown in figure~\ref{fig:1}(c) the measured $\chi^{V}_{f}$ appears to diverge near our estimated $\Gamma_{c}$. A finite size scaling analysis can be performed as well. For consistency we consider the same type of scaling form of $\chi^{V}_{f}$ as appeared in equation~(\ref{eq:chif}),

The result, plotted in figure~\ref{fig:3}(b), gives $\Gamma_{c}=0.46(2)$ and $\nu^{-1}=0.48(3)$, in fairly good agreement with what we have obtained from $U_{2}$ and $\xi$. Our final estimates are $\Gamma_{c}=0.48(1)$ and $\nu^{-1}=0.44(5)$.

\begin{figure}[!htb]
\captionsetup[subfigure]{labelformat=empty}
  \centering
  \mbox{\includegraphics[width=0.8\columnwidth]{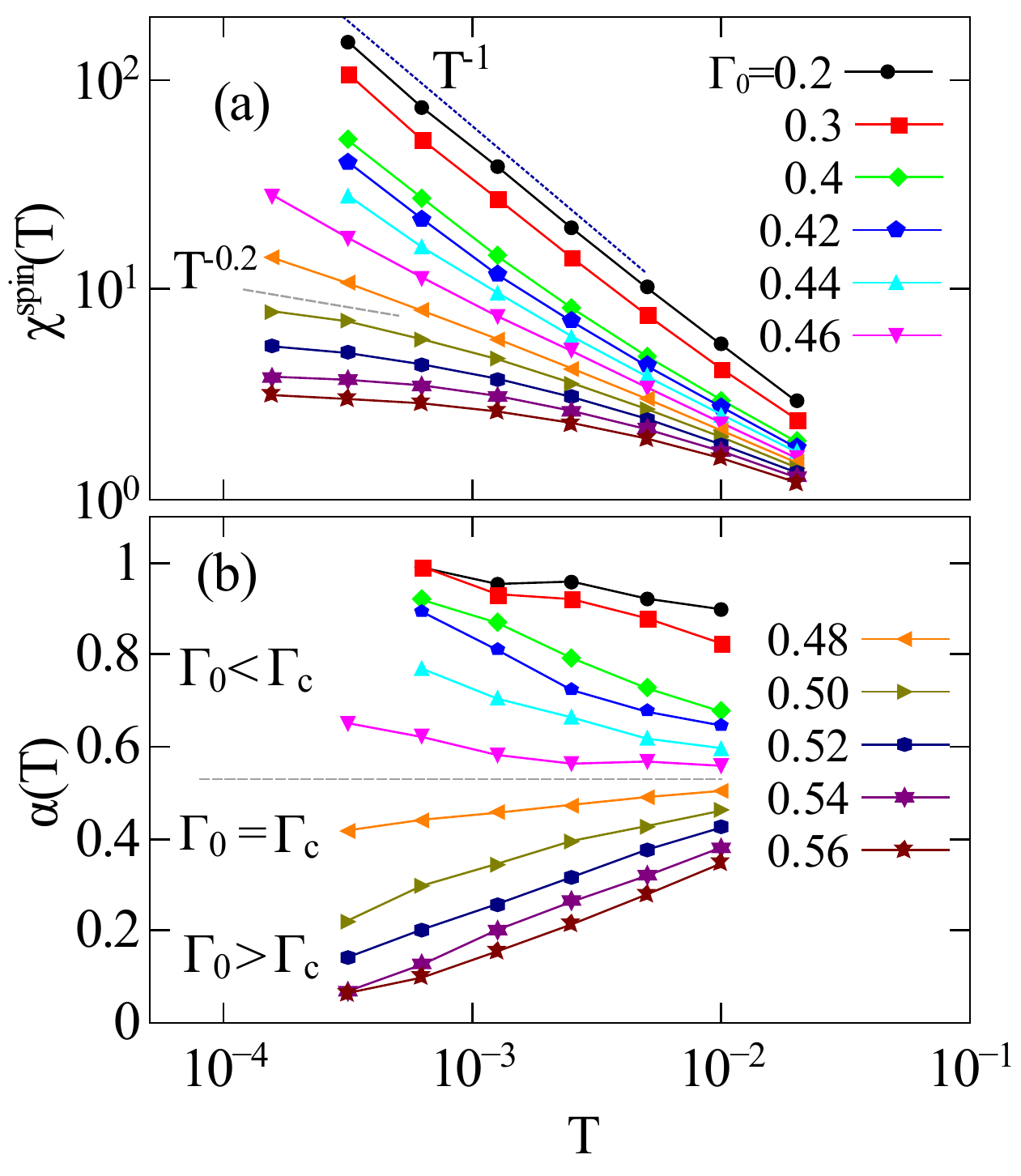}}    
\caption{(a) Temperature dependence of spin susceptibility at various $\Gamma_{0}$. Dotted and dashed lines are visual guides for the $T^{-1}$ and $T^{-s}$ behavior expected in the local moment phase and critical point C$^{\prime}$, respectively. (b) Effective power law exponent $\alpha$ (defined in text) as a function of $T$. Dashed line separates the two distinct behavior in the temperature dependence of the transient exponent $\alpha$: In the local moment regime ($\Gamma_{0}<\Gamma_{c}$) $\alpha$ will approach 1. In the Kondo regime ($\Gamma_{0}>\Gamma_{c}$) $\alpha$ will approach 0. At the QCP ($\Gamma_{0}=\Gamma_{c}$) $\alpha$ is expected to approach $0.2$.}
\label{fig:4} 
\end{figure}

After identified the location of the QCP, we now look at the temperature dependence of the spin susceptibility $\chi^{spin}$ across the QCP, shown in figure~\ref{fig:4}(a).  It turns out the critical behavior of $\chi^{spin}$ is much harder to study for the $s=0.2$ case compared to the $s=0.6$ case. For $\Gamma_{0}<\Gamma_{c}$, the dominant behavior of $\chi^{spin}(T) $ is Curie-Weiss like, reflecting the localized nature of the impurity spin. For $\Gamma_{0}>\Gamma_{c}$, $\chi^{spin}(T)$ will saturate at low T, corresponding to Kondo singlet formation. In between, we can see some indication of quantum critical behavior $\chi^{spin}(T) \propto T^{-s} $ at $\Gamma_{0}=0.50$, slightly away from our estimated $\Gamma_{c}$. We think this is due to the fact that $\chi^{spin}(T)$ at $\Gamma_{0}=\Gamma_{c}$ is still in the initial cross-over regime. To see this, we may define a transient power law exponent by $\alpha(T)=-d \log(\chi^{spin}(T) )/ d \log(T)$. For $\Gamma_{0} \leq 0.46 $ we find $\alpha(T)$ is increasing as $T$ is lowered while for $\Gamma_{0} \geq 0.48$ it is decreasing. 

We note that the calculation in reference~\cite{otsuki} has assumed the relation $\chi^{spin}(T) \propto T^{-s}$ and use it as a tool to locate the QCP by looking for the crossing point of $T^{s}\chi^{spin}(T)$ at different $T$. But there the crossing point has significant drift versus temperature, which is consistent with an evolving $\alpha(T)$ in our calculation. Here we determine the critical coupling $\Gamma_{c}$ via a variety of independent methods and obtained unambiguous results for the presence and the location of the QCP. Then we attempt to verify the critical behavior of $\chi^{spin}(T)$ directly. Unfortunately from figure~\ref{fig:4}(b) it seems that, in contrast to the case of $s=0.6$, accessing the asymptotic
critical regime requires even lower temperatures for $s=0.2$:  it appears we 
would need at least two more decades below the lowest temperature we can obtain $T=1/6400$ 
in the temperature dependence of $\chi^{spin}(T) $ to access the true critical behavior for $s=0.2$.
We interpret this as the $s=0.2$ case has a extremely low entry point $T<10^{-5}$ to reach the asymptotic 
critical regime for $\chi^{spin}(T) $, even though the bare Kondo temperature $T_{K}^{0}$ is about 
$T_{K}^{0}=2.5$ at $\Gamma=\Gamma_{c}$ in the absence of bosonic coupling. 
We have seen earlier that in the $s=0.6$ case it is much easier to access the asymptotic critical behavior 
of $\chi^{spin}(T) $.

\subsection{Phase diagram upon varying the power of the sub-ohmic spectrum}
\label{numerical.C}
The phase diagram, as specified by the two types of RG flows given in figure~\ref{fig:rg}, can be
determined for any given $0<s<1$ once we have estimated $s^{*}$. For this purpose,
we can turn to the pure bosonic problem by setting $\Gamma_{0}=0$, and vary both the bosonic coupling $g$ 
as well as the bosonic bath exponent $s$. 
As $s \rightarrow 0^{+}$, the procedure to obtain $J(\tau)$ defined in equation~(\ref{eq:mat_dos_int}) 
and equation~(\ref{eq:dos_int}) will encounter convergence issue. 
As the critical property only depends on the long time asymptotic behavior of $J(\tau)$, we directly adopt 
a $J(\tau)$ that has the correct $1/\tau^{1+s}$ dependence as our input without specifying the actual form 
of $\rho_{b}(\omega)$. To be specific, we choose $J(\tau)$ to be the following,
\begin{equation}
J(\tau)=\left[  \frac{\pi/\beta}{\sin(\pi \tau/\beta) } (1+e^{-\beta}-e^{-\tau}-e^{-(\beta-\tau)}) \right]^{1+s}.
\label{J:eq}
\end{equation}

The exponential factor will make $J(\tau)$ finite at the end points: $\lim_{\tau \rightarrow 0} J(\tau) = \lim_{\tau \rightarrow \beta} J(\tau) = 1 $. Also $J(\tau)$ is even under reflection about $\tau=\beta/2$.

We can then integrate $J(\tau)$ twice to get $B(\tau)$,
\begin{equation}
B(\tau)-B(0)=\int_{0}^{\tau} \int_{0}^{\tau^{\prime}} J(\tau^{\prime \prime}) d \tau^{\prime} d \tau^{\prime \prime} + a\tau.
\label{K:eq}
\end{equation}
with $a=-\int_{0}^{\beta/2} J(\tau^{\prime\prime}) d \tau^{\prime \prime}$ determined from the condition $d B(\tau) /d \tau |_{\tau=\beta/2} =0 $.

\begin{figure}[!htb]
\captionsetup[subfigure]{labelformat=empty}
  \centering
  \mbox{\includegraphics[width=0.9\columnwidth]{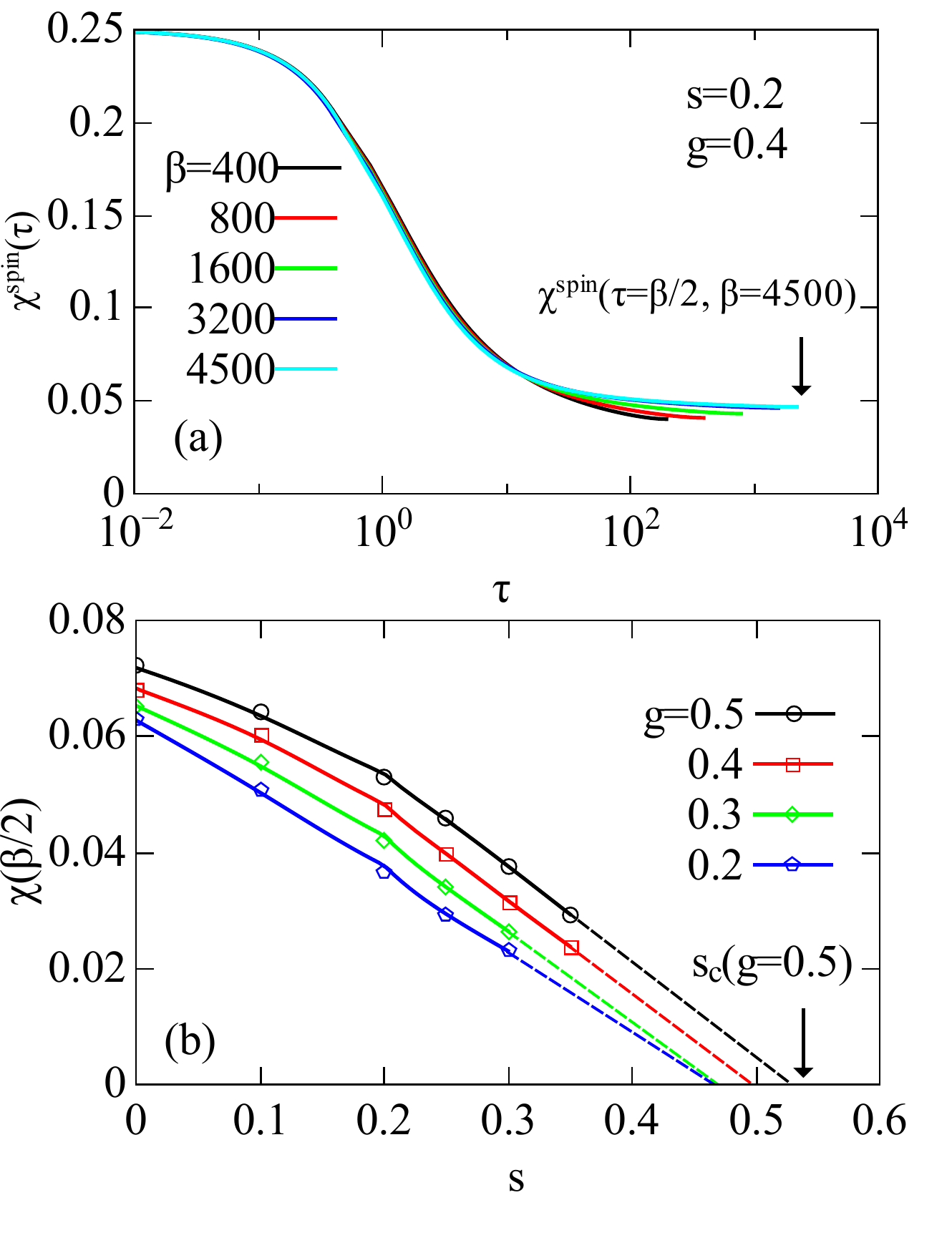}}    
  \\[-6ex]                      
\caption{(a) Dynamical spin correlation function $\chi^{s}(\tau)$ from $\tau=0$ up to $\tau=\beta/2$ at different $\beta$. For large $\beta$, $\chi^{s}(\beta/2)$ converges to a finite value around $0.05$. Arrow marks the value of  $\chi^{s}(\beta/2)$ at $\beta=4500$ (b) Effective curie constant $\chi^{s}(\beta/2)$ vs $s$ at different value of $g$. Increasing $s$ reduces the size of $\chi^{s}(\beta/2)$. Dashed lines are linear extrapolation of $\chi^{s}(\beta/2)$ to $\chi^{s}(\beta/2)=0$, the intersections with the horizontal axis give the critical values $s_{c}(g) $ for each $g$. Arrow marks the value of $s_{c}$ obtained at $g=0.5$. }
\label{chi_vs_s} 
\end{figure}

Using equation~(\ref{J:eq}) and equation~(\ref{K:eq}) as input we have obtained the dynamical spin correlation function $\chi^{s}(\tau)$ for different value of $g$ and $s$. In figure~\ref{chi_vs_s}(a) we present the result of $\chi^{s}(\tau)$ vs. $\tau$ at several different value of $\beta$ for the specific case of $g=0.4, s=0.2$. At each $\beta$, $\chi^{s}(\tau)$ drops from $1/4$ at $\tau=0$ and reaches its minimum at $\tau=\beta/2$. As $\beta$ is increased, $\chi^{s}(\tau=\beta/2)$ converge to a finite value.

\begin{figure}[!htb]
\captionsetup[subfigure]{labelformat=empty}
  \centering
   \mbox{\includegraphics[width=0.9\columnwidth]{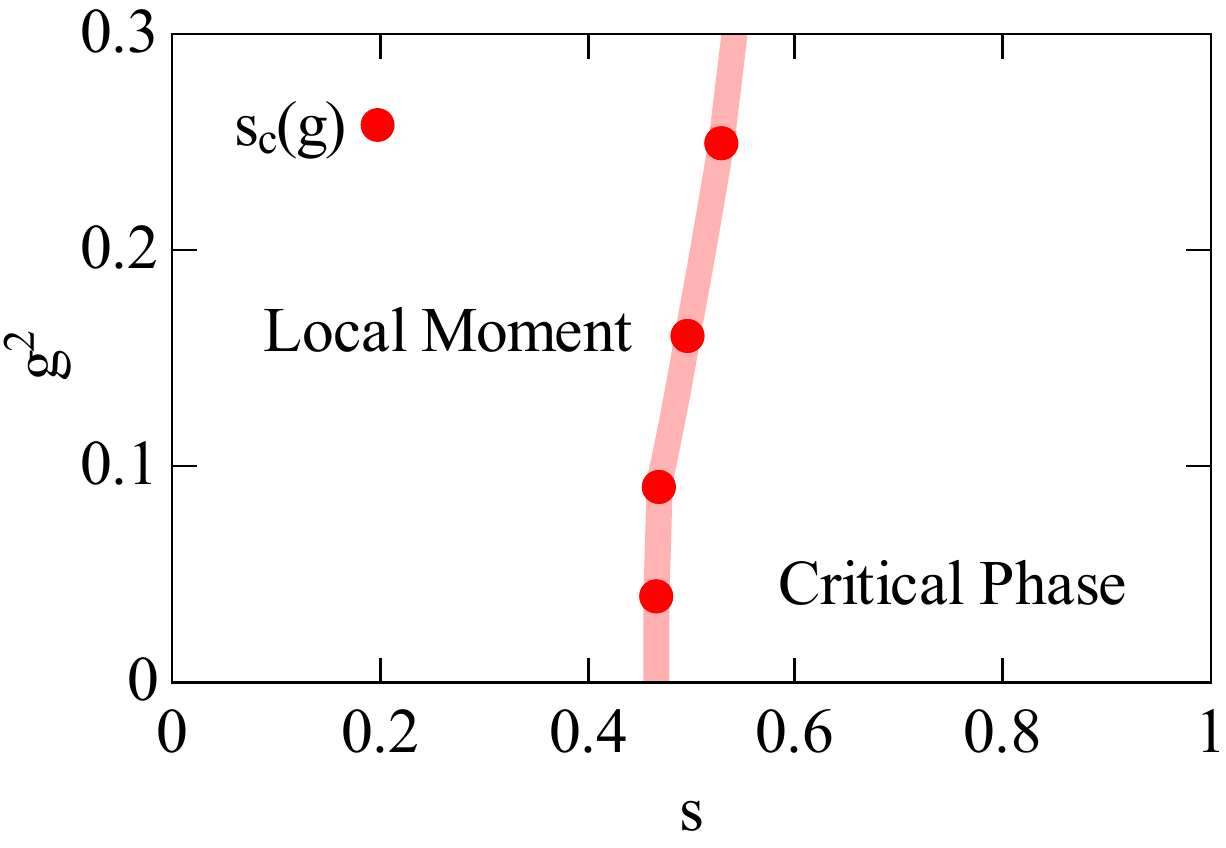}}           
\caption{  Phase diagram of the pure bosonic problem. For $s>s{*} \simeq 0.47$, increasing $g$ will induce a transition from the critical phase to the local moment phase. For $s<s{*}$, the critical phase disappears. }
\label{phase_b} 
\end{figure}

We then plot the evolution of $\chi^{s}(\tau=\beta/2)$ obtained at low temperature, as a function of $s$ for four different choices of $g$ in figure~\ref{chi_vs_s}(b), up to the smallest value of  $\chi^{s}(\tau=\beta/2)$ that we can reach convergence. We can identify $\chi^{s}(\tau=\beta/2)$ obtained here as an effective Curie constant, and use it as the order parameter for the local moment phase. We see that for fixed $g$, $\chi^{s}(\tau=\beta/2)$ decreases smoothly as a function of $s$. Furthermore, we can extrapolate each curve to larger value of $s$ until $\chi^{s}(\tau=\beta/2)$ vanishes at some critical value $s=s_{c}(g)$. This gives the value of $s$ where the corresponding $g$ is the critical value between the local moment phase and the critical phase.

The dependence of $s_{c}( g )$ on $g$  maps out the phase boundary between the local moment phase and the critical phase, which is shown in figure~\ref{phase_b}. Note that the shape of the phase boundary will depend on the specific form of $J(\tau)$ that is employed.
 As we can see the dependence of $s_{c} (g) $ on the value of $s$ is fairly weak and it reaches the $g^{2}=0$ axis at around $s=s^{*}\simeq 0.47$.  Although we do need to admit the simple extrapolation scheme performed in figure~\ref{phase_b}(a) could introduce some error. 

\section{Discussion}
\label{discussion}
Our result has important implications for the Kondo lattice model. In the EDMFT solution of the Kondo lattice model, the Kondo destruction QCP of the lattice problem is embedded in the impurity QCP of an effective BFKM. The self-consistency condition is satisfied at $s\rightarrow 0^{+}$, or $\epsilon \rightarrow 1^{-}$, provided the relation $\eta=\epsilon$ holds, which initially is an statement made at critical point C from $\epsilon$-expansion perspective. Our calculation implies that C should disappear before $\epsilon$ reaches $1$, and that the actual impurity QCP encountered in the EDFMT calculation should be C$^{\prime}$ instead. Nonetheless, the relation $\eta=\epsilon$ is still true at C$^{\prime}$, even though C and C$^{\prime}$ have different correlation length exponent, thus belonging to different universality classes. This is quite surprising until we realize that the argument that leads to $\eta=\epsilon$ only relies on the condition $\eta=\epsilon + 2 \beta(g) /g |_{g=g^{*},J=J^{*}}$, which is shown to be valid to all orders in $\epsilon$ in reference\cite{zhu2002}. The relation $\eta=\epsilon$ then follows at any intermediate coupling fixed point $g=g^{*}, J=J^{*}$, where $\beta(g) /g |_{g=g^{*},J=J^{*}}=0$, regardless of whether $g^{*}$ and $J=J^{*}$ is of the order $\epsilon$. Thereby this argument can be extended to C$^{\prime}$ as well.

\section{Conclusions}
\label{conclusions}
We have studied the SU(2) Bose-Fermi Anderson model using CT-QMC, focusing on the Kondo destruction type QCP. We find two type of such QCPs: one from Kondo screened phase to a local moment phase, the other to a critical phase. The second type QCP only exists when $s>s^{*}$, in which case the critical properties we have calculated 
agree with those from an $\epsilon$-expansion RG. At both types of QCP, our results suggest the 
spin correlation function obeys the power law $\chi^{spin}(\tau) \sim (1/\tau)^{\eta}$ with $\eta=1-s$.

\section{Acknowledgment}
{We thank H. Hu, J. Pixley and K. Ingersent for useful discussions.
The work was in part supported by the NSF (DMR-1611392), the Robert A. Welch Foundation (C-1411), 
Computing time was supported in part by the Data Analysis and Visualization Cyberinfrastructure funded by NSF under grant OCI-0959097 and an IBM Shared University Research (SUR) Award at Rice University, and by the Extreme Science and Engineering
Discovery Environment (XSEDE) by NSF under Grants No. DMR170109.
Q.S. acknowledges the hospitality  and support by a Ulam Scholarship from the Center for Nonlinear Studies at Los Alamos National Laboratory, and the hospitality of the Aspen Center for Physics, which is supported by NSF grant No. PHY-1607611.

\bibliography{SU2_1_BFKM}


\end{document}